\documentclass[11pt]{article}

\usepackage[textsize=footnotesize]{todonotes}
\usepackage{epsfig}
\usepackage{amsmath}
\usepackage{mathrsfs}
\usepackage{bm}
\usepackage{bbm}
\usepackage{amsfonts}
\usepackage{multirow}
\usepackage{booktabs}
\usepackage{setspace}
\usepackage{mathabx}
\usepackage{rotating}

\usepackage{xr}
\usepackage{array}

\usepackage[section]{placeins}
\usepackage{algorithm}
\usepackage{algorithmic}
\usepackage{graphicx,ifthen}
\usepackage{wrapfig}
\usepackage{color}
\usepackage{verbatim}
\usepackage{xcolor}
\usepackage{esint}
\usepackage{multirow}
\usepackage{xr}
\usepackage{array}
\usepackage{tabularx}
\usepackage{subfig}
\usepackage[OT1]{fontenc}
\usepackage{chngcntr}

\usepackage[bbgreekl]{mathbbol}
\DeclareSymbolFontAlphabet{\mathbbm}{bbold}
\DeclareSymbolFontAlphabet{\mathbb}{AMSb}%

\usepackage{mathtools}

\usepackage{bm}
\usepackage{natbib}
\usepackage{xr}
\usepackage{url}

\newtheorem{theorem}{Theorem}[section]
\newtheorem{lemma}{Lemma}[section]

\newtheorem{corollary}{Corollary}[section]

\newtheorem{remark}{Remark}[section]
\newtheorem{exa}{Example}[section]

\newtheorem{algo}{Algorithm}[section]
\numberwithin{equation}{section}
\numberwithin{theorem}{section}
\numberwithin{lemma}{section}
\numberwithin{proposition}{section}
\numberwithin{corollary}{section}
\numberwithin{definition}{section}
\numberwithin{cons}{section}
\numberwithin{remark}{section}
\numberwithin{exa}{section}
\numberwithin{table}{section}
\numberwithin{figure}{section}

\newcommand{\hSigma}{\widehat{\bm\Sigma}_p}
\newcommand{\tSigma}{\widetilde{\bm\Sigma}_p}

\newcommand{\bZ}{\mathbf{{Z}}}

\newcommand{\bY}{\mathbf{Y}}

\newcommand{\bE}{\mathbf{E}}

\newcommand{\bH}{\mathbf{H}}
\newcommand{\bM}{\mathbf{M}}

\newcommand{\bW}{\mathbf{W}}

\newcommand{\calC}{\mathcal{C}}
\newcommand{\calD}{\mathcal{D}}

\newcommand{\calG}{\mathcal{G}}
\newcommand{\calH}{\mathcal{H}}

\newcommand{\calK}{\mathcal{K}}
\newcommand{\calL}{\mathcal{L}}

\newcommand{\calQ}{\mathcal{Q}}

\newcommand{\calX}{\mathcal{X}}

\newcommand{\calZ}{\mathcal{Z}}
\newcommand{\bcalV}{\bm{\mathcal{V}}}

\newcommand{\frakB}{\mathfrak{B}}

\newcommand{\frakF}{\mathfrak{F}}

\newcommand{\hcalQ}{\widehat{\calQ}}
\newcommand{\tcalQ}{\widetilde{\calQ}}
\newcommand{\mE}{\mathbb{E}}

\newcommand{\mP}{\mathbb{P}}

\newcommand{\const}{\mathcal{K}}
\newcommand{\tr}{\mathrm{tr}}

\newcommand{\comp}{\mathbbm{z}}

\newcommand{\bGamma}{\calH}
\newcommand{\bR}{\mathcal{R}}
\newcommand{\bcalS}{{\mathcal{S}}}

\allowdisplaybreaks

\begin{document}

\title{High-dimensional general linear hypothesis tests \\ via non-linear spectral shrinkage\footnote{This research was partially supported by NSF grant DMS 1407530.}}

\author{
Haoran Li\footnote{Department of Statistics, University of California at Davis, One Shields Avenue, Davis,  CA 95616, USA, emails: \tt{[hrli,aaue,debpaul]@ucdavis.edu}}\ {\footnote{Corresponding author.}}
\and Alexander Aue$^\dagger$
\and Debashis Paul$^\dagger$
}
\date{\today}
\maketitle

\begin{abstract}
\setlength{\baselineskip}{1.665em}
We are interested in testing general linear hypotheses in a high-dimensional multivariate linear regression model. The framework includes many well-studied problems such as two-sample tests for equality of population means, MANOVA and others as special cases. A family of rotation-invariant tests is proposed that involves a flexible spectral shrinkage scheme applied to the sample error covariance matrix. The asymptotic normality of the test statistic under the null hypothesis is derived in the setting where dimensionality is comparable to sample sizes, assuming the existence of certain moments for the observations. The asymptotic power of the proposed test is studied under various local alternatives. The power characteristics are then utilized to propose a data-driven selection of the spectral shrinkage function. As an illustration of the general theory, we construct a family of tests involving ridge-type regularization and suggest possible extensions to more complex regularizers. A simulation study is carried out to examine the numerical performance of the proposed tests.\medskip 

%\noindent {\bf MSC 2010:} Primary: 62G99, 62H99, Secondary: 62M10, 62M15, 91B84
\end{abstract}

\setlength{\baselineskip}{1.665em}

\section{Introduction}
\label{sec:introduction}
In multivariate analysis, one of the fundamental inferential problems is to test a hypothesis involving a linear transformation of regression coefficients under a linear model. Suppose $\bY$ is a $p\times N$ matrix of observations modeled as
\begin{align}
\label{eq:linear_model}
\bY &= BX + \Sigma_p^{1/2}\bZ~,
\end{align}
where ~(i)~ $B$ is a $p\times k$ matrix of regression coefficients; ~(ii)~ $X$ is a $k\times N$ design matrix of rank $k$; ~(iii)~$\bZ$ is a $p\times N$ matrix with i.i.d.\ entries having zero mean and unit variance;  and ~(iv)~  $\Sigma_p$, a $p\times p$ nonnegative definite matrix, is the population covariance matrix of the errors, with $\Sigma_p^{1/2}$ a ``square-root'' of $\Sigma_p$ so that $\Sigma_p= \Sigma_p^{1/2}(\Sigma_p^{1/2})^T$. 
%The notation $^T$ is used to denote transposition of matrices and vectors. 
General linear hypotheses involving the linear model (\ref{eq:linear_model}) are of the form
\begin{equation}
\label{eq:general_linear_hypothesis}
H_0\colon BC=0 \qquad \mbox{vs.} \qquad H_a\colon BC\neq0,
\end{equation}
for an arbitrary $k\times q$ ``constraints matrix'' $C$, subject to the requirement that $BC$ is estimable. Without loss of generality,  $C$ is taken to be of rank $q$.  Throughout, we assume that $q$ and $k$ are fixed, even as observation dimension $p$ and sample size $N$ increase to infinity.  Henceforth, $n = N-k$ is used to denote the effective sample size, which is also the degree of freedom associated with the sample error covariance matrix.

With various choices of $X$ and $C$, the testing formulation incorporates many hypotheses of interest. For example, multivariate analysis of variance (MANOVA) is a special case. When the sample size $N$ is substantially larger than the dimension $p$ of the observations, this problem is well-studied. \citet{anderson1958introduction} and \citet{muirhead2009aspects} are among standard references. %under normality assumption of observations. %More details will be talked in Section \ref{subsec:general_linear_hypothesis_MP_law}.%Well-known statistics to test the hypothesis include Wilk's Likelyhood-ratio test, Lawley-Hotelling trace, Roy's largest root and so on, playing a central role in classical multivariate statistics.  
Various classical inferential procedures involve the matrices
\begin{align}
\hSigma 
=& \frac{1}{n} \bY(I- X^T (XX^T)^{-1} X ) \bY^T, \label{eq:Sigma_hat}\\
\widehat{\bH}_p   
=& \frac{1}{n} \bY X^T(XX^T)^{-1} C[C^T(XX^T)^{-1}C]^{-1}C^T (XX^T)^{-1} X\bY^T, \label{eq:H_hat}
\end{align}
so that $\hSigma$ is the residual covariance of the full model, an estimator of $\Sigma_p$, while $\widehat{\bH}_p$ is  the hypothesis sums of squares and cross products matrix, scaled by $n^{-1}$. In a one-way MANOVA set-up, $\hSigma$ and $\widehat{\bH}_p$ are, respectively, the within-group and between-group sums of squares and products matrices, scaled by $n^{-1}$. In the rest of the paper, we shall refer to $\hSigma$ as the sample covariance matrix. 

The testing problem (\ref{eq:general_linear_hypothesis}) is well-studied in the classical multivariate analysis literature. Three standard test procedures are the likelihood ratio test (LR), Lawley--Hotelling trace test (LH) and Bartlett--Nanda--Pillai trace (BNP) test. They are called \emph{invariant tests}, since under Gaussianity the null distributions of the test statistics are invariant with respect to $\Sigma_p$. One common feature is that all test statistics are linear functionals of the spectrum of $\widehat{\bH}_p\hSigma^{-1}$. Since this matrix is asymmetric, for convenience, a standard transformation is applied, giving the expressions of the invariant tests as follows. Define 
\begin{align}
Q_n &= X^T(XX^T)^{-1} C[C^T(XX^T)^{-1}C]^{-1/2}\label{eq:Q},\\
\bM_0 &= \frac{1}{n} Q^T_n \bY^T \hSigma^{-1} \bY Q_n. \nonumber 
\end{align}
The matrix $Q_nQ_n^T$ is the ``hat matrix'' of the reduced model under the null hypothesis. Note that the non-zero eigenvalues of $\widehat{\bH}_p\hSigma^{-1} = n^{-1}\bY Q_nQ_n^T \bY^T \hSigma^{-1}$ are the same as those of $\bM_0$. The test statistics for the LR, LH and BNP tests can be expressed as  
\begin{align*}
&T_0^{\rm LR} = \sum\nolimits_{i=1}^q \log\{1+\lambda_i(\bM_0)\}, \\
&T_0^{\rm LH} =  \sum\nolimits_{i=1}^q \lambda_i(\bM_0), \\
&T_0^{\rm BNP} = \sum\nolimits_{i=1}^q \lambda_i(\bM_0)/\{1+\lambda_i(\bM_0)\}.
\end{align*} 
The symbol $\lambda_i(\cdot)$ denotes the $i$-th largest eigenvalue of a symmetric matrix, further using the convention that $\lambda_{\max}(\cdot)$ and $\lambda_{\min}(\cdot )$ indicate the largest and smallest eigenvalue.

In contemporary statistical research and applications, high-dimensional data whose dimension is at least comparable to the sample size is ubiquitous. In this paper, focus is on the interesting boundary case when dimension and sample sizes are comparable. Primarily due to inconsistency of conventional estimators of model parameters --- such as $\hSigma$ ---, classical test procedures for the hypothesis \eqref{eq:general_linear_hypothesis} --- such as the LR, LH and BNP tests --- perform poorly in such settings. When the dimension $p$ is larger than the degree of freedom $n$, the invariant tests are not even well-defined because $\hSigma$ is singular. Even when $p$ is strictly less than $n$, but the ratio $\gamma_n = p/n$ is close to $1$, these tests are known to have poor power behavior. Asymptotic results when $\gamma_n\to \gamma \in(0,1)$ were obtained in \citet{fujikoshi2004asymptotic} under Gaussianity of the populations, and more recently in \citet{bai2017limiting} under more general settings that only require the existence of certain moments.

Pioneering work on modifying the classical solutions in high dimension is in \citet{bai2013testing}, who corrected the scaling of the LR statistic when $n\geq p$ but $p$, $k$ and $q$ are proportional to $n$. The corrected LR statistic was shown to have significantly more power than its classical counterpart. In contrast, in this paper, we focus on the setting where $k$ and $q$ are fixed even as $n,p \to \infty$ so that $\gamma_n=p/n\to \gamma\in(0,\infty)$. In the multivariate regression problem, this corresponds to a situation where the response is high-dimensional, while the predictor is finite-dimensional. In the MANOVA problem, this framework corresponds to high-dimensional observations belonging to one of a finite number of populations.

To the best of our knowledge, when $n<p$, the linear hypothesis testing problem has been studied in depth only for specific submodels of \eqref{eq:linear_model}, primarily for the important case of two-sample tests for equality of population means.
%Recently, many existing methodologies for the two sample problem are generalized to one-way MANOVA and other more general model designs. 
%In view of the fact that some of the core ideas behind two-sample tests in high dimensions have been generalized to MANOVA and  other linear hypothesis testing problems, here we give a brief overview of the growing literature on the two-sample mean testing problem. 
For the latter tests, a widely used idea is to construct modified statistics based on replacing $\hSigma^{-1}$ with an appropriate substitute. This approach was pioneered in \citet{bai1996effect} and further developed in %considered $A = I_p$. Later, 
\citet{chen2010two}. % removed some technical limitations of the test by \citet{bai1996effect}  through a clever modification of the test statistic, that extended its applicability even to the $p/n\to \infty$ setting.  Both these methods avoid the scaling of $\widehat{\bH}_p$ by the observation covariance structure and therefore bypass the estimation of $\Sigma_p$. This main theme has been applied in different forms
Various extensions to %the problem 
one-way MANOVA \citep{srivastava2006multivariate, yamada2015testing, srivastava2006multivariate, hu2017testing} and a general multi-sample Behrens--Fisher problem under heteroscedasticity \citep{zhou2017high} exist.
Other notable works for the two-sample problem include \citet{biswas2014nonparametric, chang2017simulation,chen2014two,guo2016tests,lopes2011more,srivastava2016raptt,wang2015high}.
A second approach aims to regularize $\hSigma$ to address the issue of its near-singularity in high dimensions; see \citet{chen2011regularized} and \citet{LiAuePaulPengWang} for ridge-type penalties in two-sample settings. %seek to regularize $\hSigma$ by adding a ridge-type penalty to the sample covariance matrix, which corresponds to specifying $A_{\lambda} = (\hSigma +\lambda I_p)^{-1}$, for $\lambda >0$, as the scaling matrix, instead of $\hSigma^{-1}$ in the formulation of the Hotelling's $T^2$ test. This leads to a rotation-invariant test. 
Finally, another alternative line of attack consists of exploiting sparsity; see \citet{tony2014two, cai2014high}. %, who assumed a form of sparsity for the difference of mean vectors. Assuming further that a ``good'' estimate of the precision matrix is available, they constructed tests based on the maximum component-wise mean difference of suitably transformed observations. The idea was extended by \citet{cai2014high} to one-way MANOVA.

In  this paper, we seek to regularize the spectrum of $\hSigma$ by flexible shrinkage functions. 
%investigate testing the hypothesis (\ref{eq:general_linear_hypothesis}) with flexible non-linear spectral shrinkage. 
For a symmetric $p\times p$ matrix $A$ and a function $g(\cdot)$ on $\mathbb{R}$, define 
\[
g(A) = R_A \mbox{diag}\big(g(\lambda_1(A)),\dots, g(\lambda_p(A))  \big) R_A^T,
\] 
where $R_A$ is the matrix of eigenvectors associated with the ordered eigenvalues of $A$.
Now, consider any real-valued function $f(\cdot)$ on $\mathbb{R}$ that is analytic over a specific domain associated with the limiting behavior of the eigenvalues of $\hSigma$, as elaborated in Section \ref{sec:methodology}. The proposed statistics are functionals of eigenvalues of the regularized quadratic forms
\[\bM(f) = \frac{1}{n} Q_n^T \bY^T f(\hSigma) \bY Q_n.\]
Specifically, we propose regularized versions of LR, LH and BNP test criteria, namely 
\begin{align*} 
T^{\rm LR}(f)  &= \sum\nolimits_{i=1}^q \log\{1+\lambda_i(\bM(f))\},\\
T^{\rm LH}(f)  &= \sum\nolimits_{i=1}^q \lambda_i(\bM(f)),\\
T^{\rm BNP}(f)&= \sum\nolimits_{i=1}^q \lambda_i(\bM(f))/\{1+\lambda_i(\bM(f))\}. 
\end{align*}
These test statistics are designed to capture possible departures from the null hypothesis, when $\hSigma$ is replaced by $f(\hSigma)$, while suitable choices of the regularizer $f$ allow for getting around the problem of singularity or near-singularity when $p$ is comparable to $n$.

Notice that $\bM(f)$ has the same non-zero eigenvalues as $f(\hSigma)\widehat{\bH}_p$. Thus, the proposed test family is a generalization of the classical statistics based on $\hSigma^{-1} \widehat{\bH}_p$. Importantly, $\bM(f)$ --- and consequently the proposed statistics --- is \emph{rotation-invariant}, which means if a linear transformation is applied to the observations with an arbitrary orthogonal matrix, the statistic remains unchanged. It is a desirable property when not much additional knowledge about $\Sigma_p$ and $BC$ is available. %for example, sparsity or a graph model.
%We will work under the scenario $p/n\to \gamma \in (0,\infty)$, %assuming $k$ and $q$ are fixed. 
It should be noted that the two-sample mean tests by \citet{bai1996effect} and \citet{LiAuePaulPengWang}, together with their generalization to MANOVA, are special cases of the proposed family with $f(x) = 1$ and $f(x) = 1/(x+\lambda)$, $\lambda>0$, respectively.

The present work builds on the work by \citet{LiAuePaulPengWang}. The theoretical analysis also involves an extension of the analytical framework adopted by \citet{pan2011central} in their study of the asymptotic behavior of Hotelling's $T^2$ statistic for non-Gaussian observations. However, the current work goes well beyond the existing literature in several aspects. We highlight these as the key contributions of this manuscript: (a) We propose new families of rotation-invariant tests for general linear hypotheses for multivariate regression problems involving high-dimensional response and fixed-dimensional predictor variables that incorporate a flexible regularization scheme to account for the dimensionality of the observations growing proportional to the sample size. (b) Unlike \citet{LiAuePaulPengWang}, who assumed sub-Gaussianity, here only the existence of finite fourth moments of the observations is required. (c) Unlike  \citet{pan2011central}, who assumed $\Sigma_p = I_p$, $\Sigma_p$ is allowed to be fairly arbitrary and subjected only to some standard conditions on the limiting behavior of its spectrum. 
%\textcolor{magenta}{Our theoretical contributions are elaborated in the Appendix.} 
(d) We carry out a detailed analysis of the power characteristics of the 
proposed tests. The proposal of %under the general linear hypothesis problem, 
a class of local alternatives enables a clear interpretation of the contributions of different parameters in the performance of the test. (e) We develop a data-driven test procedure based on the principle of maximizing asymptotic power under appropriate local alternatives. This principle leads to the definition of a composite test that combines the optimal tests associated with a set of different kinds of local alternatives. The latter formulation is an extension of the data-adaptive test procedure designed by \citet{LiAuePaulPengWang} for the two-sample testing problem.

The rest of the paper is organized as follows. Section \ref{sec:methodology} introduces the asymptotics of the proposed test family both under the null hypothesis and under a class of local alternatives. Using these local alternatives, in Section \ref{sec:locally_most_powerful_test} a data-driven shrinkage selection methodology based on maximizing asymptotic power is developed.
% under the class of local alternatives introduced in Section \ref{sec:methodology}. 
In Section \ref{sec:ridge_regu_higher_order}, an application of the asymptotic theory and the shrinkage selection method is given for the ridge-regularization family. An extension of ridge-regularization to higher orders is also discussed. The results of a simulation study are reported in Section \ref{sec:simulation}. In the Appendix, a proof outline of the main theorem is presented, while technical details and proofs of other theorems are collected in the Supplementary Material, which is available at \url{anson.ucdavis.edu/%7Elihaoran/}.

\section{Asymptotic theory}
\label{sec:methodology}

After giving necessary preliminaries on \emph{Random Matrix Theory} (RMT), the asymptotic theory of the proposed tests under the null hypothesis and under various local alternative models is presented in this section. For any $p\times p$ symmetric matrix $A$, define the \emph{Empirical Spectral Distribution} (ESD) $F^A$ of $A$ by
\[F^A(\tau) = p^{-1} \sum\nolimits_{i=1}^p \mathbbm{1}_{\{\lambda_i(A)\leq \tau\} }.\]
In the following, %$\|\cdot\|_\infty$ means the supermum absolute value of a function over its domain and 
$\|\cdot\|_{\max}$ stands for %the matrix element-wise maximum ``norm'', that is 
the maximum absolute value of the entries of a matrix. The following assumptions are employed.
\begin{itemize}\itemsep0.3em
\item[\textbf{C1}] 
(Moment conditions) $\bZ$ has i.i.d. entries $z_{ij}$ such that
$\mE z_{ij}=0$, $\mE z_{ij}^2 = 1$, $\mE z_{ij}^4 <\infty$;
\item[\textbf{C2}]
(High-dimensional setting) $k$ and $q$ are fixed, while $p,n\to\infty$ such that $\gamma_n = p/n \to \gamma \in (0,\infty)$ and $\sqrt{n}|\gamma_n-\gamma|\to0$; 
\item[\textbf{C3}]
(Boundedness of spectral norm) $\Sigma_p$ is non-negative definite; $\limsup_p\lambda_{\max}(\Sigma_p)<\infty$; 
\item[\textbf{C4}]
(Asymptotic stability of ESD) There exists a distribution $L^{\Sigma}$ with compact support in $[0, \infty)$, non-degenerate at zero, such that $\sqrt{n} {\mathcal D}_{W}(F^{\Sigma_p}, L^{\Sigma}) \to 0$, as $n,p\to\infty$, where $\calD_{W}(F_1, F_2)$ denotes the \emph{Wasserstein distance} between distributions $F_1$ and $F_2$, defined as 
\[\calD_{W}(F_1,F_2) =  \sup_f\Big\{ \Big|\int fdF_1 - \int f dF_2 \Big|\colon f \mbox{ is 1-Lipschitz} \Big\}.\]
\item[\textbf{C5}]
(Asymptotically full rank) $X$ is of full rank and $n^{-1}XX^T$ converges to a positive definite $k\times k$ matrix. Moreover, $\limsup_{n\to\infty}\|X\|_{\max}<\infty$;
\item[\textbf{C6}]
(Asymptotically estimable)  $\liminf_{n\to\infty}\lambda_{\min}(C^T(n^{-1}XX^T)^{-1}C)>0$. 
\end{itemize}

\subsection{Preliminaries on random matrix theory}
\label{subsec:background}

%In RMT, the limiting behavior of the spectrum of $ N^{-1}\sum_{j=1}^N \Sigma_p^{1/2}\bZ \bZ^T (\Sigma_p^{1/2})^T$ is well-studied. The matrix, typically unobservable in practice, is an uncentered version of $\hSigma$. The distinction between $\hSigma$ and the uncentered version is not material for mostly all of the interesting results in RMT, especially when one is only concerned with the limiting ESD. It is due to the fact that $n^{-1}\Sigma_p^{1/2}\bZ X^T(XX^T)^{-1}X\bZ^T(\Sigma_p^{1/2})^T$ is of rank $k$ and the rank inequality (see Theorem A.43 of \citet{bai2010spectral}). We then have the cornerstone of the paper as follows, obtained by modifying results in RMT regarding the spectrum of the uncentered sample covariance matrix. 

Recall that the Stieltjes transform $m_G(\cdot)$ of any function $G$ of bounded variation on $\mathbb{R}$ is defined by 
\[
m_G(\mathfrak{\comp}) = \int_{-\infty}^\infty \frac{dG(x)}{x-\comp}, 
\qquad \comp\in \mathbb{C}^+ \coloneqq \{u+iv\colon v>0\}.
\]
Minor modifications of a standard RMT result imply that, under Conditions \textbf{C1}--\textbf{C6}, the ESD $F^{\hSigma}$ converges almost surely to a nonrandom distribution $F^\infty$ at all points of continuity of $F^\infty$. This limit is determined in such a way that for any $z\in\mathbb{C}^{+}$, the Stieltjes transform $m(\cdot)=m_{F^\infty}(\cdot)$ of $F^\infty$ is the unique solution in $\mathbb{C}^{+}$ of the equation 
\begin{equation}
m(\comp)= \int \frac{dL^{\Sigma}(\tau)}{\tau(1-\gamma-\gamma \comp m(\comp))-\comp}.
\label{eq:M-P-Eq}
\end{equation}
Equation (\ref{eq:M-P-Eq}) is often referred to as the Mar\v{c}enko--Pastur equation. 
Moreover, pointwise almost surely for $\comp\in \mathbb{C}^+$, $m_{F^{\hSigma}}(\comp)$ converges to $m_{F^\infty}(\comp)$. The convergence holds even when $\comp\in \mathbb{R}_{-}$ (negative reals) with a smooth extension of $m_{F^\infty}$ to $\mathbb{R}_{-}$. 
Readers may refer to \citet{%silverstein1995strong,silverstein1995analysis,
bai2004clt} and \citet{paul2014random} %,LiAuePaulPengWang
for more details.  From now on, for notational simplicity, we shall write  $m_{F^\infty}(\comp)$  as $m(\comp)$ and write $m_{F^{\hSigma}}(\comp)$ as  $m_{n,p}(\comp)$. Note that
\[
m_{n,p}(\comp) = p^{-1}\tr (\hSigma-\comp I_p)^{-1}
\] 
and 
%Up to now, the importance of the resolvent $(\hSigma- \comp I_p)^{-1}$ presents itself. 
define
\begin{equation}\label{eq:Theta}
\Theta(\comp,\gamma)= \{1-\gamma -\gamma\comp m(\comp)\}^{-1}.
\end{equation}
% Apart from its trace, 
It is known that $(\hSigma- \comp I_p)^{-1}$, for any fixed $\comp\in \mathbb{C}^+$, has a \emph{deterministic equivalent} (\citet{bai2004clt,liu2015marvcenko,LiAuePaulPengWang}),  given by
\[\{\Theta^{-1}(\comp,\gamma) \Sigma_p -\comp I\}^{-1}, \] 
in the sense that for symmetric matrices $A$ bounded in operator norm, as $n\to \infty$,
\[p^{-1} \tr[(\hSigma -\comp I_p)^{-1}A] - p^{-1} \tr[\{\Theta^{-1}(\comp,\gamma) \Sigma_p -\comp I\}^{-1}A] \to 0, \mbox{~~~with probability 1.}\]
Resolvent and deterministic equivalent will be used frequently in this paper. They will appear for example as Cauchy kernels in contour integrals in various places.

\subsection{Asymptotics under the null hypothesis}
\label{subsec:asym_null}
%In this subsection, we present the asymptotics under the null hypothesis of $M(f)$ and the proposed statistics. 
To begin with, for $k\geq 1$, denote by $\bW = [w_{ij}]_{i,j=1}^k$ the \emph{Gaussian Orthogonal Ensemble} (\textbf{GOE}) defined by (1)~ $w_{ij}=w_{ji}$;  (2) $w_{ii} \sim \mathcal{N}(0,1)$, $w_{ij}\sim \mathcal{N}(0,1/2)$, $i\neq j$; (3) $w_{ij}$'s are jointly independent for $1 \leq i \leq j \leq k$.   
%
%\textcolor{red}{
%Should we extend f to be any holomorphic function and re-formulate the theorem? 
  %Keep in mind, compared to ARHT paper, $\Theta^{new} = 1+ \gamma \Theta_1^{old}$,  $\delta^{new} = \gamma \Theta_2^{old}$
  %the definition of GOE changed, Oct 11. 
%}
%
Throughout this paper, $f(\cdot)$ is assumed to be analytic in an open interval containing 
\[
\calX \coloneqq [0,~\limsup\nolimits_{p\to\infty}\lambda_{\max}(\Sigma_p)(1+\sqrt{\gamma})^2].
\] 
Let $\calC$ to be a closed contour enclosing $\calX$ such that $f(\cdot)$ has a complex extension to the interior of $\calC$. Further use $\calC^2$ to denote $\calC \otimes \calC = \{(\comp_1,\comp_2)\colon \comp_1, \comp_2\in\calC\}$. %The asymptotic null distribution is determined in the next theorem.

\begin{theorem}
Suppose \textbf{C1}--\textbf{C6} hold. Under the null hypothesis $H_0\colon BC=0$,
\[ {\sqrt{n}}\{\bM(f) - \Omega(f,\gamma) I_q \} \Longrightarrow {\Delta^{1/2}(f,\gamma)} \bW, \]
where $\Longrightarrow$ denotes weak convergence and  $\Omega(f,\gamma)$ and $\Delta(f,\gamma)$ are as follows. 
\[\Omega(f,\gamma) = \frac{-1}{2\pi i}\oint_\calC f(\comp) (\Theta(\comp,\gamma)-1) d\comp.\]
See (\ref{eq:Theta}) for the definition of $\Theta(\comp,\gamma)$. For any two analytic functions $f_1$ and $f_2$, 
\begin{equation*}
\Delta(f_1,f_2,\gamma) = \frac{2}{(2\pi i)^2}\oiint_{\calC^2} f_1(\comp_1)f_2(\comp_2) \delta(\comp_1,\comp_2,\gamma)d\comp_1 d\comp_2,
\end{equation*}
and $\Delta(f,f,\gamma)$ is written as $\Delta(f,\gamma)$ for simplicity. The kernel $\delta(\comp_1,\comp_2,\gamma)$ is such that
\begin{align*}
\delta(\comp_1,\comp_2,\gamma) 
&=  {\Theta(\comp_1,\gamma)\Theta(\comp_2,\gamma)} \Big[\frac{\comp_1 \Theta(\comp_1,\gamma) - \comp_2 \Theta(\comp_2,\gamma)}{\comp_1-\comp_2} -1\Big],\\
\delta(\comp,\comp,\gamma) 
&= \lim_{\comp_2\to\comp} \delta(\comp,\comp_2,\gamma) = {\Theta^2(\comp,\gamma)}\Big[ \frac{\partial \comp \Theta(\comp,\gamma)}{\partial\comp} -1\Big] \\
& = \gamma\{1+ \comp m(\comp)\} \Theta^3(\comp,\gamma) + \gamma\comp \{ m(\comp) +\comp m'(\comp)\}\Theta^4(\comp,\gamma).
\end{align*}
The contour integral is taken counter-clockwise.
\label{thm:asmp_null}
\end{theorem}

%Following from Theorem \ref{thm:asmp_null}, with 
Using knowledge of the eigenvalues of the \textbf{GOE} leads to the following statement.
\begin{corollary}
Under the conditions of Theorem \ref{thm:asmp_null}, assume further that $\Delta(f,\gamma)> 0$. Let 
\[
\tilde{\lambda}_i =\frac{\sqrt{n}}{ \Delta^{1/2}(f,\gamma)} \{\lambda_i(\bM(f)) - \Omega(f,\gamma)\}, 
\qquad i = 1,\dots,q. 
\]
Then, the limiting joint density function of $(\tilde{\lambda}_1,\dots,\tilde{\lambda}_q)$ at $y_1\geq y_2\geq \cdots \geq y_q $ is given by 
\[\Big(2^{q/2} \prod_{i=1}^q \Gamma(i/2)\Big)^{-1}\prod_{i<j}(y_i-y_j) \exp\Big(-\frac{1}{2}\sum\limits_{i=1}^q y_i^2\Big). \] 
\label{corollary:eigenvalues_GOE}
\end{corollary}

Although without closed forms, $\Omega(f,\gamma)$ and $\Delta(f,\gamma)$ do not depend on the choice of $\calC$ used to compute the contour integral. 
%form is actually common in RMT literature. It appears because the usage of \emph{Cauchy's integral formula} in the proof of the theorem.  
With the resolvent as kernel $\bM(f)$ can be expressed as the integral of  $f(\comp) n^{-1} Q_n^T \bY^T (\hSigma -\comp I_p)^{-1} \bY Q_n$ on any contour $\calC$, up to a scaling factor. The quadratic form $n^{-1} Q_n^T \bY^T (\hSigma -\comp I_p)^{-1} \bY Q_n$ is then shown to concentrate around $[\Theta(\comp,\gamma)-1] I_q$, which consequently serves as the integral kernel in $\Omega(f,\gamma)$. The kernel $\delta(\comp_1,\comp_2,\gamma)$ of $\Delta(f,\gamma)$ is the limit of $\mE[n^{-1}\tr\{(\hSigma -\comp_1 I_p)^{-1}\Sigma_p(\hSigma -\comp_2 I_p)^{-1}\Sigma_p\}]$. 

%The asymptotic mean and variance of $\bM(f)$ involve contour integrals on $\calC$, not available in closed forms. Similar formulae can be found in RMT literature, for instance \citet{bai2004clt}. The integrals appear because we use Cauchy's integral formula to prove the theorem. Under some conditions,  $\bM(f)$ is expressed as the integral of $f(\comp) n^{-1} Q_n^T \bY^T (\hSigma -\comp I_p)^{-1} \bY Q_n$. The kernel $n^{-1} Q_n^T \bY^T (\hSigma -\comp I_p)^{-1} \bY Q_n$ is shown to be asymptoticly normal.  
%The kernel $\Theta(\comp,\gamma)-1$ and $\delta(\comp_1,\comp_2,\gamma)$ in $\Omega(f,\gamma)$ and $\Delta(f,\gamma)$ are 

%The integral kernel in the asymptotic mean $\Omega(f,\gamma)$ is 
%\[\Theta(\comp,\gamma)-1 = \lim_{n,p\to\infty} n^{-1}\tr[(\Theta^{-1}(\comp,\gamma) \Sigma_p -\comp I)^{-1}\Sigma_p)]\]
%under \textbf{C1-C6}. The kernel $\delta(\comp_1,\comp_2,\gamma)$ in the asymptotic variance $\Delta(f,\gamma)$ is also closely related to $(\hSigma -\comp I_p)^{-1}$ and $[\Theta^{-1}(\comp,\gamma) \Sigma_p -\comp I]^{-1}$. Indeed, under \textbf{C1-C6},
%\[\delta(\comp_1,\comp_2,\gamma)= \lim_{n,p\to\infty}\mE n^{-1}\tr[(\hSigma -\comp I_p)^{-1}\Sigma_p(\hSigma -\comp I_p)^{-1}\Sigma_p].\]

\begin{remark}
\label{remark:0_variance}
Two sufficient conditions for $\Delta(f,\gamma)>0$ are 
\vspace{-.2cm}
\begin{itemize} \itemsep-.5ex
\item[(1)] $f(x)>0 \mbox{ for } x \in \calX$;
\item[(2)] $f(x)\geq 0 \mbox{ for } x \in \calX$, with $f(x)\neq0$ for some $x\in\calX$, and  $\liminf\lambda_{\min}(\Sigma_p) >0.$
\end{itemize}
\end{remark}

It would be convenient if $\Omega(f,\gamma)$ and $\Delta(f,\gamma)$ had closed forms in order to avoid computational inefficiencies. Closed forms are available for special cases as shown in the following lemma. 
%Because in practice, to approximate the integral numerically is computationally inefficient.  We have the following results for a special case of $f$. 
\begin{lemma}
\label{lemma:special_form_ARHT}
%In Theorem \ref{thm:asmp_null}, when $f_{a}(x) = (x-a)^{-1}$, with $a\notin \mathbb{R}^+ \cup \{0\}$, $\Omega(f_a, \gamma)$ and $\Delta(f_a,\gamma)$ have the following explicit form. 
%\begin{align*}
%&\Omega(f_{a},\gamma) = \Theta(a,\gamma) -1 = \frac{\gamma + \gamma a m(a)}{1- \gamma -\gamma a m(a)}, \\[5pt]
%&\Delta(f_{a},\gamma) =  \delta(a,a,\gamma) =\gamma\{1+ a m(a)\} \Theta^3(a,\gamma) + \gamma a \{ m(a) + a m'(a)\}\Theta^4(a,\gamma). 
%\end{align*}
When $f(x,\ell) = (x-\ell)^{-1}$ with $\ell\in \mathbb{R}^-$, %entries of $\bM(f)$ are also asymptoticly (complex) normal. It can be verified by applying Theorem \ref{thm:asmp_null} to any linear combination of the real part and the imaginary part of $f$. 
the contour integrals in Theorem \ref{thm:asmp_null} have closed forms, namely, for $j$, $j_1$, $j_2$ $= 0,1,2,\dots$,
\begin{align*}
&\frac{-1}{2\pi i}\oint_\calC \frac{\partial^j f(\comp,\ell)}{\partial \ell^j} (\Theta(\comp,\gamma)-1)d\comp  = \frac{\partial^{j}(\Theta(\ell,\gamma) -1)}{\partial \ell^{j}},\\[5pt] %= \frac{\gamma + \gamma a m(a)}{1- \gamma -\gamma a m(a)}, 
& \frac{1}{(2\pi i)^2}  \oiint_{\calC^2}\frac{\partial^{j_1} f(\comp_1,\ell_1)}{\partial \ell^{j_1}_1} \frac{\partial^{j_2} f(\comp_2,\ell_2)}{\partial \ell^{j_2}_2} \delta(\comp_1,\comp_2, \gamma) d\comp_1 d\comp_2  = \frac{\partial^{j_1+j_2}\delta(\ell_1,\ell_2,\gamma)}{\partial \ell^{j_1}_1 \partial \ell^{j_2}_2}.
\end{align*}
The results continue to hold when $\ell \in \mathbb{C}\setminus\calX$. 
\end{lemma}
Lemma \ref{lemma:special_form_ARHT} indicates that it is possible to have convenient and accurate estimators of the asymptotic mean and variance of $\bM(f)$ under ridge-regularization. The result easily generalizes to the setting when $f(x)$ is a linear combination of functions of the form $(x-\ell_j)^{-1}$, for any finite collection of $\ell_j$'s. We elaborate on this in Section \ref{sec:ridge_regu_higher_order}.
%Note, the forms coincide with their counterparts in \citet{chen2011regularized} and \citet{LiAuePaulPengWang}.

To conduct the tests, consistent estimators of $\Omega(f,\gamma)$ and $\Delta(f,\gamma)$ are needed.
\begin{lemma}
\label{lemma:emp_counterpart}
Let $\widehat{\Theta}(\comp,\gamma_n)$ and $\widehat{\delta}(\comp_1,\comp_2,\gamma_n)$ be the plug-in estimators of ${\Theta}(\comp,\gamma)$ and ${\delta}(\comp_1,\break\comp_2,\gamma)$, with $(m(\comp),\gamma)$ estimated by $(m_{n,p}(\comp),\gamma_n)$. For general $f$, $f_1$, $f_2$, we can estimate $\Omega(f,\gamma)$ and $\Delta(f_1,f_2,\gamma)$ by replacing ${\Theta}(\comp,\gamma)$ and $\delta(\comp_1,\comp_2,\gamma)$ with $\widehat{\Theta}(\comp,\gamma_n)$ and $\widehat{\delta}(\comp_1,\comp_2,\gamma_n)$. Denote the resulting estimators by $\widehat{\Omega}(f,\gamma_n)$ and $\widehat{\Delta}(f_1,f_2,\gamma_n)$.  Then, 
\begin{align*}
&\sqrt{n} |\widehat{\Omega}(f,\gamma_n) -\Omega(f,\gamma) | \stackrel{P}{\longrightarrow}0, \\[5pt]
&\sqrt{n} |\widehat{\Delta}(f_1,f_2,\gamma_n) - \Delta(f_1,f_2,\gamma)| \stackrel{P}{\longrightarrow}0, 
\end{align*}
where $\stackrel{P}{\longrightarrow}$ indicates convergence in probability. Again, we write $\widehat{\Delta}(f,f,\gamma_n)$ as $\widehat{\Delta}(f,\gamma_n)$. %It implies Theorem \ref{thm:asmp_null} will also hold with these empirical counterparts plugged in.  

For the special case of $f^{(j)}(x,\ell) =\partial^j (x-\ell)^{-1}/\partial\ell^j$, $j=0,1,2,\dots$ and $\ell\in \mathbb{C}\setminus \calX$, using Lemma \ref{lemma:special_form_ARHT}, natural estimators in closed forms are 
\begin{align*}
&\widehat{\Omega}(f^{(j)}(x,\ell) ,\gamma_n)= \frac{\partial^{j}(\widehat\Theta(\ell,\gamma_n) -1)}{\partial \ell^{j}}, \\
&\widehat{\Delta}(f^{(j_1)}(x,\ell_1), f^{(j_2)}(x,\ell_2),\gamma_n) = \frac{\partial^{j_1+j_2}2\widehat\delta(\ell_1,\ell_2,\gamma_n)}{\partial \ell^{j_1}_1 \partial \ell^{j_2}_2}. 
\end{align*}
In particular, for $j,j_1,j_2=0$, 
\begin{align*}
&\widehat{\Omega}(f(x,\ell),\gamma_n) =  \widehat\Theta(\ell,\gamma_n) -1,\\
&\widehat{\Delta}(f(x,\ell_1),f(x,\ell_2),\gamma_n)= 2\widehat{\delta}(\ell_1,\ell_2,\gamma_n).
\end{align*}
The estimators are consistent, for any fixed $j$ and $\ell$. Given the eigenvalues of $\hSigma$, the computational complexity of calculating the above estimators is $O(p)$.
%\begin{align*}
%\sqrt{n} &\Big|\frac{\partial^{j}(\widehat\Theta(\ell,\gamma_n) -1)}{\partial \ell^{j}} - \frac{\partial^{j}(\Theta(\ell,\gamma) -1)}{\partial \ell^{j}}\Big| \stackrel{P}\longrightarrow 0,\\[5pt]
%& \frac{\partial^{j_1+j_2}\widehat\delta(\ell_1,\ell_2,\gamma)}{\partial \ell^{j_1}_1 \partial \ell^{j_2}_2} -\frac{\partial^{j_1+j_2}\delta(\ell_1,\ell_2,\gamma)}{\partial \ell^{j_1}_1 \partial \ell^{j_2}_2}\stackrel{P}\longrightarrow 0. 
%\end{align*} 
\end{lemma}

Recall the definitions of $T^{\rm LR}(f), T^{\rm LH}(f)$ and $T^{\rm BNP}(f)$ from Section \ref{sec:introduction}.
%Theorem \ref{thm:asy_null_three_stat} follows immediately from Theorem \ref{thm:asmp_null}. 
\begin{theorem}
\label{thm:asy_null_three_stat}
Suppose \textbf{C1}--\textbf{C6} hold and $\Delta(f,\gamma)>0$. Under $H_0\colon BC=0$, 
\begin{align*}
\widehat{T}^{\rm LR}(f) 
&\coloneqq\frac{\sqrt{n}\{1+\widehat{\Omega}(f,\gamma_n)\} }{q^{1/2}\widehat{\Delta}^{1/2}(f,\gamma_n)} [T^{\rm LR}(f) - q \log\{1+\widehat{\Omega}(f,\gamma_n)\}] 
{\Longrightarrow} \mathcal{N}(0,1),\\  
\widehat{T}^{\rm LH}(f) 
&\coloneqq\frac{\sqrt{n}}{q^{1/2}\widehat{\Delta}^{1/2}(f,\gamma_n)}\{T^{\rm LH}(f) - q\widehat{\Omega}(f,\gamma_n)\}{\Longrightarrow} \mathcal{N}(0,1),\\  
\widehat{T}^{\rm BNP}(f) 
&\coloneqq\frac{\sqrt{n}\{1+\widehat{\Omega}(f,\gamma_n)\}^2}{q^{1/2}\widehat{\Delta}^{1/2}(f,\gamma_n)}\Big\{T^{\rm BNP}(f) - q\frac{\widehat{\Omega}(f,\gamma_n)}{1+\widehat{\Omega}(f,\gamma_n)}\Big\} {\Longrightarrow} \mathcal{N}(0,1). 
\end{align*}
%Denote $\xi_\alpha$ to be the $1-\alpha$ quantile of standard normal. 
For any of the three tests, the null hypothesis is rejected at asymptotic level $\alpha$, if \break$\widehat{T}(f) > \xi_\alpha$, where $\xi_\alpha$ is the $1-\alpha$ quantile of the standard normal distribution.
\end{theorem} 

%\begin{remark}
%Theorem \ref{thm:asy_null_three_stat} still holds if we only require $\mbox{Re}(f(x))\geq0$ on $\mR^+\cup\{0\}$. %In that case, we may reject the null hypothesis at level $\alpha$, if  $|\widehat{T}(f)| > \xi_\alpha$. 
%\label{remark:complex_f_statistics}
%\end{remark}

\subsection{Asymptotic power under local alternatives}
\label{subsec:asym_power}

This subsection deals with the behavior of the proposed family of tests under a host of local alternatives. We start with deterministic alternatives, a framework commonly used in the literature to study the asymptotic power of inferential procedures. Next, we consider a Bayesian framework, using a class of priors that characterize the structure of the alternatives. Because the results to follow simultaneously hold for $\widehat{T}^{\rm LR}(f)$, $\widehat{T}^{\rm LH}(f)$ and $\widehat{T}^{\rm BNP}(f)$, the unifying notation $\widehat{T}(f)$ will be used to refer to each of the test statistics.

\subsubsection{Deterministic local alternatives}

Consider a sequence of $BC$ such that, on an open subset of $\mathbb{C}$ containing $\calX$,
\begin{equation}
\sqrt{n} C^T B^T \{\Theta^{-1}(\comp,\gamma) \Sigma_p - \comp I\}^{-1} BC \longrightarrow D(\comp,\gamma)\quad \mbox{ pointwise, as }n,p\to\infty.
\label{eq:determ_local_alternatives}
\end{equation} 
Observe that $\bY Q_n =\sqrt{n} BC [C^T (n^{-1} XX^T )^{-1} C]^{-1/2} + \Sigma_p^{1/2}\bZ Q_n$ and define
\begin{equation}
\bGamma(D,f) = T^{-1/2} \Big[\frac{-1}{2\pi i} \oint_\calC f(\comp) D(\comp,\gamma)  d\comp\Big]T^{-1/2}, \mbox{ where}
\label{eq:Omega_def}
\end{equation}
\begin{equation}\label{eq:T_matrix_def}
T = \lim_{n\to\infty} C^T (n^{-1} XX^T )^{-1} C. 
\end{equation}
Note that $T$ exists and is non-singular under \textbf{C5} and \textbf{C6}. {If further $f(x)\geq 0$ for any $x \in \calX$, $\bGamma (D,f)$ is non-negative definite. 
}
\begin{theorem}
\label{thm:local_power_Mf}
Suppose \textbf{C1}--\textbf{C6} and (\ref{eq:determ_local_alternatives}) hold, and $\Delta(f,\gamma)>0$. Then, as $n\to\infty$,
\[
\frac{\sqrt{n}}{\Delta^{1/2}(f,\gamma)} \{\bM(f) - \Omega(f,\gamma) I_q \}   {\Longrightarrow} \bW + \frac{\bGamma(D,f)}{\Delta^{1/2}(f,\gamma)}.
\]
\end{theorem}

Denote the power functions of $\widehat{T}(f)$ at asymptotic level $\alpha$, conditional on $BC$, by  
 %\begin{align*}
 %&\Upsilon_{LR}(BC,f) = \mathbb{P}(\widehat{T}_{LR}(f) >\xi_\alpha \mid BC),\\
 %&\Upsilon_{LH}(BC,f) = \mathbb{P}(\widehat{T}_{LH}(f) >\xi_\alpha \mid BC),\\ 
 %&\Upsilon_{BNP}(BC,f) = \mathbb{P}(\widehat{T}_{BNP}(f) >\xi_\alpha \mid BC). 
 %\label{eq:power_function }
 %\end{align*}
 \[
 \Upsilon(BC,f) = \mathbb{P}(\widehat{T}(f) >\xi_\alpha \mid BC).
 \]
The asymptotic behavior of the power functions is described in the following corollary. %of Theorem \ref{thm:local_power_Mf}.   
\begin{corollary}
Under the assumptions of Theorem~\ref{thm:local_power_Mf}, as $n\to\infty$, %Suppose \textbf{C1}--\textbf{C6} and (\ref{eq:determ_local_alternatives}) hold and $\Delta(f,\gamma)>0$. 
\[
\Upsilon(BC,f) \longrightarrow \Phi\Big(-\xi_\alpha + \frac{\tr(\bGamma(D,f))}{q^{1/2} \Delta^{1/2}(f,\gamma)}\Big), \]
where $\Phi$ is the standard normal CDF. 
\label{corollary:local_power_three_stat} 
\end{corollary}

\begin{remark}
Corollary \ref{corollary:local_power_three_stat} indicates the three proposed statistics have identical asymptotic powers under the assumed local alternatives. This is because the first-order Taylor expansions of $x$, $\log(1+x)$ and $x/(1+x)$ coincide at $0$. However, the respective empirical powers may differ considerably for moderate sample sizes. %depending on different types of alternatives. 
\label{remark:local_power_identical}
\end{remark}

The following remark provides a sufficient condition under which \eqref{eq:determ_local_alternatives} is satisfied.
Denoting the columns of $BC$ by $[\mu_1,\dots,\mu_q]$, it follows that
\[\sqrt{n} C^T B^T \{\Theta^{-1}(\comp,\gamma) \Sigma_p - \comp I\}^{-1} BC = \sqrt{n} \Big[\mu_i^T \{\Theta^{-1}(\comp,\gamma)\Sigma_p -\comp I_p\}^{-1}\mu_j \Big]_{i,j=1}^q. \]

\vspace{-5pt}
\begin{remark}
(a) Let $\bE_{m,p}$ denote the eigen-projection associated with $\lambda_{m,p}=\lambda_m(\Sigma_p)$. Suppose that there exists a sequence (in $p$) of mappings $[\frakB_{ij;p}]_{i,j=1}^q$ from $[0,\infty)^{q^2} $ to $[0,\infty)^{q^2}$, satisfying  $\frakB_{ij;p}(\lambda_{m,p}) = \sqrt{n} p \mu_i^T \bE_{m,p} \mu_j$, $ m =1,\dots,p$, and a mapping\break $[\frakB_{ij;\infty}]_{i,j=1}^q$ continuous on $[0,\infty)^{q^2}$ such that, as $p\to\infty$ and for $1\leq i, j \leq q$,
\[
\int |\frakB_{ij;p}(x) - \frakB_{ij;\infty}(x)| d F^{\Sigma_p}(x) \to 0.
\]
Then, under \textbf{C4}, it follows that (\ref{eq:determ_local_alternatives}) holds with $D(\comp,\gamma) = [d_{ij}(\comp,\gamma)]_{i,j=1}^q$ and
\[d_{ij}(\comp,\gamma)=\int \frac{\frakB_{ij;\infty}(x) d L^{\Sigma}(x)}{x\Theta^{-1}(\comp,\gamma) -\comp}  =\int \frac{\frakB_{ij;\infty}(x) d L^{\Sigma}(x)}{x\{ 1-\gamma - \gamma \comp m(\comp) \} -\comp}. \]

(b) If $\Sigma_p = I_p$, then (\ref{eq:determ_local_alternatives}) is satisfied if $\sqrt{n} \mu_i^T\mu_j \to \const_{ij}$, for some constants $\const_{ij}$, $1\leq i, j \leq q$. In this case, $D(\comp,\gamma) =(\Theta^{-1}(\comp,\gamma) -\comp)^{-1} [\const_{ij}]_{i,j=1}^q$. 
\label{remark:project_mu_eigenSigma}
\end{remark}

\subsubsection{Probabilistic local alternatives}
\label{subsubsec:probabilistic_local_alternatives}

While deterministic local alternatives provide useful information, they are somewhat restrictive for the purpose of a systematic investigation of the power characteristics. Therefore, probabilistic alternatives are considered in the form of a sequence of prior distributions for $BC$. This has the added advantage of providing flexibility for incorporating structural information about the regression parameters and the constraints matrices.
%The framework is not an innovation of our paper, but first formulated in \citet{LiAuePaulPengWang}. 
%Under the framework, we can incorporate structure of $BC$, usually %reasonable to assume when pre-knowledge of $BC$ is available. 
%In many fields such as spatio-temporal statistics, pre-knowledge of $BC$ is potentially available and we might have specific   % based on potentially available pre-knowledge of the alternatives.
%
The proposed formulation of probabilistic alternatives can be seen as an extension of the proposal adopted by \citet{LiAuePaulPengWang} in the context of two-sample tests for  equality of means. One challenge associated with formulating meaningful alternatives to the hypothesis \eqref{eq:general_linear_hypothesis}, when compared to the two-sample testing problem, is that there are many more plausible ways in which the null hypothesis can be violated. Considering this, we propose a class of alternatives, that on one hand can incorporate a multitude of structures of the parameter $BC$, while on the other hand retains analytical tractability in terms of providing interpretable expressions for the local asymptotic power.

Assume the following prior model of $BC$ with separable covariance
\begin{equation}
BC = n^{-1/4} p^{-1/2} \bR \bcalV \bcalS^T,
\label{eq:proba_local_alter_model}
\end{equation}
where $\bcalV$ is a $p\times m$ stochastic matrix ($m\geq1$ fixed) with independent elements $\nu_{ij}$ such that $\mE[\nu_{ij}] =0$, $\mE[|\nu_{ij}|^2] =1$ and $\max_{ij}\mE[ |\nu_{ij}|^4] \leq p^{c_\nu} $ for some $c_{\nu} \in (0,1)$; $\bR$ is a $p\times p$ deterministic matrix and $\bcalS$ is a fixed $q\times m$ matrix. Moreover,  let $\|\bR\|_2 \leq \calK_1<\infty$ and suppose there is a nonrandom function $h(\comp,\gamma)$ such that, as $p\to\infty$, on an open subset of $\mathbb{C}$ containing $\calX$,
\begin{equation}
p^{-1} \tr\{(\Theta^{-1}(\comp,\gamma) \Sigma_p - \comp I)^{-1} \bR\bR^T\} \to h(\comp,\gamma)
\qquad\mbox{ pointwise}. 
\label{eq:covariance_prior_condition}
\end{equation} 
Recalling that $(\Theta^{-1}(\comp,\gamma) \Sigma_p - \comp I)^{-1}$ is the deterministic equivalent of the resolvent $(\hSigma - \comp I)^{-1}$, existence of the limit
\eqref{eq:covariance_prior_condition} also implies that 
$p^{-1} \tr\{(\hSigma - \comp I)^{-1}\bR\bR^T\}$ converges pointwise in probability to $h(\comp,\gamma)$. Notice also that $p^{-1} \tr\{(\hSigma - \comp I)^{-1}\bR\bR^T\}$ is the Stieltjes transform of a measure supported on the eigenvalues of $\hSigma$.

Model \eqref{eq:proba_local_alter_model} leads to a fairly broad covariance design for multi-dimensional random elements, encompassing structures commonly encountered in many application domains, especially in spatio-temporal statistics. We give some representative examples by considering various functional forms of the matrix $\bcalS$. Denote by $\mu_j$ the columns of $BC$ and by $V_j$ the columns of $\bcalV$.
\begin{exa}
\label{ex}
{\rm 
In all that follows $j$ takes values in $1,\ldots,q$. 
\vspace{-.2cm}
\begin{itemize}\setlength\itemsep{-0.1em}
%\item \textbf{Model 1} (identical): $\mu_j = n^{-1/4} p^{-1/2} \bR \nu_0$, $j=1,\dots,q$,  where $\bR$ is a $p\times p$ matrix and $\nu_0$ is a random vectors with independent coordinates $\nu_{0j}$'s such that $\mE[\nu_{0j}] =0$, $\mE[|\nu_{0j}|^2] =1$ and $\max_j\mE |\nu_{0j}|^4 \leq p^{c_\nu} $ for some $c_{\nu} \in (0,1)$. Moreover,  let $\|\bR\| \leq \calK_1<\infty$, and, as $n,p\to\infty$,

%Under the model, (\ref{eq:prob_local_alternative}) is reached with $D(\comp,\gamma) = h(\comp,\gamma) \frakS$, where $\frakS = \mathbf{1}_q \mathbf{1}_q^T$.

\item[(a)] {\it Independent}: $\mu_j = n^{-1/4} p^{-1/2} \bR V_j$; %, $j=1,\dots,q$;%, where $\{\nu_j\}_{j=1}^q$ are columns of $\bcalV$ and $\bcalS = I_q$.
%Under the model, (\ref{eq:prob_local_alternative}) is reached with $D(\comp,\gamma) = h(\comp,\gamma) \frakS$, where $\frakS = I_q$.

\item[(b)] {\it Longitudinal}: $\mu_j = n^{-1/4} p^{-1/2} \bR(V_1 + V_2 j + \cdots + V_m j^{m-1})$; %$j=1,\dots,q$;%,  where $\bR$ and $\nu_0$ are specified as in Model 1 and $\nu_1$ is independent and identically distributed as $\nu_0$.  
%Under the model, (\ref{eq:prob_local_alternative}) is reached with $D(\comp,\gamma) = h(\comp,\gamma) \frakS$, where $\frakS = \Big[1+ij\Big]_{i,j=1}^q$.

%\item \textbf{Model 4} (AR(1)): $\mu_j = \phi_1 \mu_{j-1} + n^{-1/4} p^{-1/2} \bR\nu_j$, with $|\phi_1|<1$,  $\bR$ and $\{\nu_j\}_{j=1}^q$ are specified as in Model 3. 
%Under the model, (\ref{eq:prob_local_alternative}) is reached with $D(\comp,\gamma) = h(\comp,\gamma) \frakS$, where $\frakS = \Big[ {\phi_1^{|i-j|}}/{(1-\phi_1^2)}\Big]_{i,j=1}^q$.

\item[(c)] {\it Moving average}: $\mu_j = n^{-1/4} p^{-1/2} \bR [V_{j+t} + \theta_1V_{j+t-1} +\cdots + \theta_tV_{j}]$ for constants $\theta_1,\ldots,\theta_t$. %and $\bR$, $\nu_j$ are specified as in Model 1. Under the model, (\ref{eq:prob_local_alternative}) is reached with $D(\comp,\gamma) = h(\comp,\gamma) \frakS$, where $\frakS = I_q + \Big[\theta_1^{\max\{0,2-|i-j|\}} \Big]_{i,j=1}^q $.
\end{itemize}
}
\end{exa}

Taking the MANOVA problem to illustrate, suppose that the columns of $B$ represent group mean vectors, and suppose $C$ is the matrix that determines successive contrasts among them. % mean vectors. 
Then, $\mu_j$ is the difference between the means of group $j$ and group $j+1$. 
Parts (a)--(c) of Example \ref{ex} correspond then to $\mu_1,\ldots,\mu_q$ respectively following an independent, a longitudinal and a moving average process. The row-wise covariance structure is assumed to be such that each $\mu_j$ has a covariance matrix proportional to $n^{-1/2}p^{-1}\bR\bR^T$. The factor $n^{-1/2}p^{-1}$ provides the scaling for the tests to have non-trivial local power. %$\bR$ is assumed to satisfy (\ref{eq:covariance_prior_condition}). 

A sufficient condition that leads to (\ref{eq:covariance_prior_condition}), similar to Remark \ref{remark:project_mu_eigenSigma}, is to postulate the existence of functions $\tilde{\frakB}_p$ satisfying $\tilde{\frakB}_p(\lambda_{j,p}) = \tr\{\bE_{j,p} \bR\bR^T\}$, $j=1,\dots,p$, and 
\[\int |\tilde{\frakB}_p(x) -\tilde{\frakB}_{\infty}(x)|d F^{\Sigma_p}(x) \to 0\] 
for some function $\tilde{\frakB}_\infty$ continuous on $[0,\infty)$, where $\lambda_{j,p}$ is the $j$th eigenvalue of $\Sigma_p$ and  $\bE_{j,p}$ is the eigen-projection associated with $\lambda_{j,p}$. Then %$h(\comp,\gamma)$ will have the form 
\begin{equation}
 h(\comp,\gamma) = \int\frac{\tilde{\frakB}_\infty(x) dL^{\Sigma}(x)}{x\{1-\gamma-\gamma\comp m(\comp)\}-\comp} .
\label{eq:h_form}
\end{equation}
Equations \eqref{eq:covariance_prior_condition} and \eqref{eq:h_form} indicate that $h(\comp,\gamma)$ effectively captures the distribution of the total spectral mass of $\bR\bR^T$ across the spectral coordinates of $\Sigma_p$, also taking into account the dimensionality effect through the aspect ratio $\gamma$. Later, we shall discuss specific classes of the matrices $\bR$ that lead to analytically tractable expressions for $h(\comp,\gamma)$, with the structure of $\bR$ linking the parameter $BC$ under the alternative through \eqref{eq:proba_local_alter_model} to the structure of $\Sigma_p$.

Another important feature of the probabilistic model is that it incorporates both dense and sparse alternatives through different specifications of the innovation variables $\nu_{ij}$. 
We consider two special cases.
\begin{enumerate}\itemsep-.1ex
\item \textit{Dense alternative:}  $\nu_{ij} \sim \mathcal{N}(0,1)$;
\item \textit{Sparse alternative:} $\nu_{ij} \sim G_\eta$, for some $\eta\in(0,1)$, where $G_\eta$ is the discrete probability distribution assigning mass $1-p^{-\eta}$ to $0$ and mass $(1/2)p^{-\eta}$ to the points $\pm p^{\eta/2}$. 
\end{enumerate}
Note that the usual notion of sparsity corresponds to the setting
where in addition, $\bR = I_p$. More generally, the second specification above formulates a prior model for $BC$ that is sparse
in the coordinate system determined by $\bR$. In particular, if $\bR\bR^T$ is a polynomial in $\Sigma_p$ (see Section~\ref{subsec:polynomial_alternatives} for a discussion), $BC$ can be seen as sparse in the spectral coordinates of $\Sigma_p$.

%Recalling the definition of $\bGamma(D,f)$ in (\ref{eq:Omega_def}),  observe that under each of the models,
%\[\tr(\bGamma(D,f)) =  \Big(\frac{-1}{2\pi i}\oint_{\calC}f(\comp) h(\comp,\gamma) d\comp\Big)  \tr(\frakS T^{-1}).  \]

\begin{theorem}
Suppose that \textbf{C1}--\textbf{C6} hold and $\Delta(f,\gamma)>0$. Also suppose that, under $H_a$, $BC$ has a prior distribution given by (\ref{eq:proba_local_alter_model}). Then,
the power function of each of the three test statistics satisfies
\begin{equation}
\Upsilon(BC,f) \stackrel{L_1}\longrightarrow \Phi\Big(-\xi_\alpha + \frac{ \tr(\bcalS\bcalS^T T^{-1}) }{ q^{1/2} \Delta^{1/2}(f,\gamma)}\oint_{\calC} \frac{-1}{2\pi i}f(\comp) h(\comp,\gamma) d\comp\Big) ,
\label{eq:converge_local_power_prob}
 \end{equation}
as $n\to\infty$, where 
%$\Upsilon$ denotes any of $\Upsilon_{LR}$, $\Upsilon_{LH}$ or $\Upsilon_{BNP}$, $\Phi$ is the standard normal CDF, and 
$T$ is as in (\ref{eq:T_matrix_def}) and {$\stackrel{L_1}\longrightarrow$ indicates $L_1$-convergence (with respect to the prior measure of $BC$)}.
\label{thm:probability_local_power_three_stat}
\end{theorem}

\begin{remark}
Even if the quantity $h_p(\comp,\gamma) = p^{-1} \tr\{( \Theta^{-1}(\comp,\gamma) \Sigma_p - \comp I)^{-1} \bR\bR^T\}$ does not converge, it can be verified that the difference between the left- and right-hand sides of \eqref{eq:converge_local_power_prob} still converges to zero in $L_1$ if $h(\comp,\gamma)$ is replaced by $h_p(\comp,\gamma)$.
\label{remark:h_not_converge}
\end{remark}

Observe that the matrices $\bR$ and $\bcalS$ decouple in the expression (\ref{eq:converge_local_power_prob}) for the asymptotic power. Dependence on the unknown error covariance matrix $\Sigma_p$, besides $\Delta^{1/2}(f,\gamma)$, is only through the 
function $h(\comp,\nu)$, which incorporates the structure of the matrix $\bR\bR^T$.  It is also noticeable that distributional characteristics of the variables $\nu_{ij}$ do not affect the asymptotic power. Indeed, the proposed tests have the same local  asymptotic power under both sparse and dense alternatives.

%The theorem indicates the power factor is also separable  $\tr(\bcalS\bcalS^T T^{-1})$, $ \Delta^{-1/2}(f,\gamma)$ and $\oint f(\comp) h(\comp,\gamma) d\comp$. The effect of correlation structure between columns of $BC$ on the asymptotic power reflects in term of $\tr(\frakS T^{-1})$.  The effect of coordinate-wise covariance is represented by $\oint f(\comp) h(\comp,\gamma) d\comp$. $\Delta(f,\gamma)$ is a scaling factor.

\section{Data-driven selection of shrinkage}
\label{sec:locally_most_powerful_test}

In this section, we introduce a data-driven procedure to select the ``optimal'' $f$ from a parametric family $\frakF$ of shrinkage functions.  The strategy is to maximize the local power function $\Upsilon(BC,f)$ over $f$, given a class of probabilistic local alternatives as in \eqref{eq:proba_local_alter_model}. In designing the classes of alternatives, we focus our attention only on the specification of $\bR$. This is because, as the expression \eqref{eq:converge_local_power_prob} shows, the dependence on the matrix $\bcalS$ is only through a multiplier involving a ``known'' matrix $T$, while the effect of the unknown covariance $\Sigma_p$ (and its interaction with $\bR$) manifests itself through the function $h(\comp,\gamma)$. Another reason for focusing on $\bR$ is that the choice of $\bcalS$ is closely related to the specific type of linear model being considered, while the choice of $\bR$ is associated with the structure of the error distribution.

We present some settings of $BC$ for which $h(\comp,\gamma)$ can be computed explicitly. We also verify that the standardized test statistic with the data-driven selection of $f$ is still asymptotically standard normal under suitable conditions. Hence, the Type 1 error rate of the tests is asymptotically not inflated, although the same data is used for both shrinkage selection and testing. Lastly, we present a composite test procedure that combines the optimal tests corresponding to different prior models of $BC$ and thereby improves adaptivity to various kinds of alternatives.

\subsection{Shrinkage family}
\label{subsec:shrinkage_family}

Suppose the family of shrinkage functions is such that 
\[\frakF = \{f_{\ell}\colon \ell \in  \calL\}, \]
%$\frakF$ is a family of analytic functions $f_{\ell}$ indexed by a parameter $\ell\in \calL$ such that 
\begin{itemize}\itemsep-.1ex
\item[(i)] $\calL$ is a compact subset of $\mathbb{R}^r$, $r\in \mathbb{N}^+$;
\item[(ii)] There is a closed, connected subset $\calZ$ of $\mathbb{C}$ such that  %$\calX \subset \calZ$, recalling that 
$\calX= [0,~\limsup_p\lambda_{\max}(\Sigma_p)(1+\sqrt{\gamma})^2]\subset \calZ$, and the third-order partial derivatives of $f_{\ell}$ with respect to $\ell$ are continuous on $\calL\otimes \calZ$;
\item[(iii)] The gradient  $\nabla_{\ell}f_{\ell}$ and the Hessian $\nabla^2_{\ell} f_{\ell}$ of $f_{\ell}$ with respect to $\ell$ have analytic extensions to $\calZ$ for all $\ell\in \calL$;
%\item[(ii)] The third-order partial derivatives $\nabla^3_{\ell}f_{\ell}$ of $f_{\ell}$ with respect to $\ell$ exist and are bounded on $\calL$;
%\item[(iii)] We can find a open region in $\mathbb{C}$ containing $[0,\limsup \lambda_{\max}(\Sigma_p)(1+\sqrt{\gamma})^2  ]$ such that  $f_{\ell}$, $\nabla_{\ell} f_{\ell}$ and $\nabla^2_{\ell} f_{\ell}$  have complex analytic extensions to the region, for any fixed $\ell\in \calL$;
\item[(iv)] $\inf_{\ell\in \calL}\Delta(f_\ell,\gamma)>0$.
\end{itemize}
Under the probabilistic prior model \eqref{eq:proba_local_alter_model} with $h(\comp,\gamma)$ in (\ref{eq:covariance_prior_condition}) given, define 
\[\Xi(\ell,h,\gamma) = \frac{-1}{2\pi i \Delta^{1/2}(f_\ell,\gamma)} \oint_{\calC}  f_\ell(\comp) h(\comp,\gamma) d\comp.\]
Theorem \ref{thm:probability_local_power_three_stat} suggests that $\ell$ should be chosen such that $\Xi(\ell,h,\gamma)$ is maximized, that is,
\[
\ell_{opt} = \mbox{arg}\max_{\ell\in \calL}\Xi(\ell,h,\gamma).
\]
The test with the selected shrinkage will then be the locally most powerful test under the alternatives specified by \eqref{eq:proba_local_alter_model} and \eqref{eq:covariance_prior_condition} for any given choice of $\bcalS$. Since $\Xi(\ell,h,\gamma)$ is continuous with respect to $\ell$ under condition (i)--(iv), $\ell_{opt}$ exists. Importantly, $\Xi(\ell,h,\gamma)$ does not rely on $\bcalS$. In other words, different column-wise covariance structures of $BC$ are uniform in terms of selecting the optimal shrinkage. This significantly simplifies the selection procedure. 

Recall that $h(\comp,\ell)$ is the limit of $p^{-1}\tr\{(\Theta^{-1}(\comp,\gamma)\Sigma_p -\comp I)^{-1}\bR\bR^T\}$.  We next present two possible settings of $\bR\bR^T$ under which $h(\comp,\gamma)$ and consequently $\Xi(\ell,h,\gamma)$ can be accurately estimated: %Two possible scenarios are considered. 
\begin{itemize}\itemsep-.1ex
\item[(1)] Suppose $\bR\bR^T$ is specified. Then, $h(\comp,\gamma)$ is estimated by 
$\widehat{h}(\comp,\gamma_n) = p^{-1}\tr\{(\hSigma-\comp I)^{-1}\bR\bR^T\}$ and
\[
\widehat{\Xi}(\ell,\widehat{h},\gamma_n) \coloneqq \frac{-1}{2\pi i \widehat\Delta^{1/2}(f_\ell,\gamma_n)} \oint_{\calC}  f_{\ell}(\comp) \widehat{h}(\comp,\gamma_n) d\comp
\]
is a consistent estimator of $\Xi(f,h,\gamma)$. As an example of this scenario, assume that the $p$ components of $\mu_j$ admit a natural ordering such that the dependence between their coordinates is a function of the difference between their indexes. Then we may set $\bR\bR^T$ to be a Toeplitz matrix (stationary auto-covariance structure).

\item[(2)] Only  the spectral mass distribution of $\bR\bR^T$ in the form of $\tilde{\frakB}_\infty$ described in \eqref{eq:h_form} is specified. 
\end{itemize}
The remainder of this section is devoted to dealing with the second scenario.

\subsection{Polynomial alternatives}
\label{subsec:polynomial_alternatives}

Even if $\tilde{\frakB}_\infty$ is given, the estimation of $h(\comp,\gamma)$ is still challenging since it involves the unknown limiting spectral distribution $L^{\Sigma}$. In order to estimate $h(\comp,\gamma)$, it is convenient to have it in a closed form. It is feasible if $\tilde{\frakB}_\infty$ is a polynomial, which is true if $\bR\bR^T$ is a matrix polynomial in $\Sigma_p$. Since any smooth function can be approximated by polynomials, this formulation is quite flexible and practically beneficial. Assume therefore that 
\begin{equation}
\label{eq:poly_approx_bRbR}
\bR\bR^T  = \sum\nolimits_{j=0}^s {t}_j \Sigma_p^j,
\end{equation} 
where $t_0,\dots,t_s$ are pre-specified weights such that $\sum_{j=0}^s {t}_j \Sigma^j_p$ is nonnegative definite. 
Under the model, 
\[ 
h(\comp,\gamma)
= \lim_{p\to\infty} p^{-1} \tr[(\Theta^{-1}(\comp,\gamma)\Sigma_p -\comp I )^{-1} \sum\nolimits_{j=0}^s {t}_j \Sigma_p^j] 
=  \sum\nolimits_{j=0}^s t_j \rho_j(\comp,\gamma),
\]
where the functions $\rho_j(\comp,\gamma)$ satisfy the recursive formula \citep[see][]{ledoit2011eigenvectors}
\[ 
\rho_0(\comp,\gamma) = m(\comp),
\qquad \rho_{j+1}(\comp,\gamma) = \Theta(\comp,\gamma) \Big[ \int x^j dL^{\Sigma}(x) + \comp \rho_j(\comp,\gamma)\Big].
\] 
For any $j\in \mathbb{N}$, $\int x^j dL^{\Sigma}(x)$, and consequently $\rho_j(\comp,\gamma)$, can be estimated consistently \citep[Lemma 1]{bai2010estimation}. Specifically, $p^{-1}\tr(\hSigma)$ is a consistent estimator of $\int x dL^{\Sigma}(x)$. 

In practice, we restrict to the case $s = 2$. There are several considerations that
guided this choice of $s$ as stated in \citet{LiAuePaulPengWang}. First, for $s = 2$, all quantities involved can be computed explicitly without requiring knowledge of higher-order moments of the observations. Also, the corresponding estimating equations for $h(\comp,\gamma)$ are more stable as they do not involve higher-order spectral moments. Second, the choice of $s = 2$ yields a
significant, yet nontrivial, concentration of the prior covariance of $\mu_j$, $j=1,\dots,q$, (that is $\bR\bR^T$ up to a scaling factor) in the directions of the leading eigenvectors of $\Sigma_p$. Finally, the choice $s=2$ allows for both convex and concave shapes of the spectral mass distribution $\tilde{\frakB}_\infty$ since the latter becomes a quadratic function.

With $s=2$, we estimate $\rho_0(\comp,\gamma)$, $\rho_1(\comp,\gamma)$, $\rho_2(\comp,\gamma)$, and $h(\comp,\gamma)$ by
\begin{equation}
\label{eq:rho0_3}
\begin{split}
&\widehat{\rho}_0(\comp,\gamma_n) = m_{n,p}(\comp), \\
&\widehat{\rho}_1(\comp,\gamma_n) = \widehat{\Theta}(\comp,\gamma_n)[1 + \comp m_{n,p}(\comp)], \\ 
&\widehat{\rho}_2(\comp,\gamma_n) = \widehat{\Theta}(\comp,\gamma_n)\big[p^{-1}\tr(\hSigma) + \comp \widehat{\rho}_1(\comp,\gamma_n) \big],\\
&\widehat{h}(\comp,\gamma_n) = \sum\nolimits_{j=0}^2 t_j \widehat{\rho}_j(\comp,\gamma_n).
\end{split}
\end{equation} 

The algorithm for the data-driven shrinkage selection is stated next. 
\begin{algo}[Data-driven shrinkage selection] {~} \vspace{-.2cm}
\begin{itemize}\itemsep-.2ex
\item[1.] Specify prior weights $\tilde{t} = (t_0,t_1,t_2)$. The canonical choices are $(1,0,0)$, $(0,1,0)$, $(0,0,1)$;
\item[2.] Compute $\widehat{h}(\comp,\gamma_n) = \sum_{j=0}^2 t_j \widehat{\rho}_j(\comp,\gamma_n)$;
\item[3.] For any $\ell\in \calL$, numerically compute the integral 
\[\widehat{\Xi}(\ell,\widehat{h},\gamma_n) = \frac{-1}{2\pi i \widehat\Delta^{1/2}(f_\ell,\gamma_n)} \oint_{\calC}  f_{\ell}(\comp) \widehat{h}(\comp,\gamma_n) d\comp ;\]
\item[4.] Select $\ell_{opt}(\tilde{t}) = \mbox{arg}\max_{\ell\in \calL}\widehat{\Xi}(\ell,\widehat{h},\gamma_n)$.  
\end{itemize}
\label{algo:empricial_selection_shrinkage}
\end{algo}
The behavior of the tests applied with the data-driven shrinkage selection is described in the following theorem. 
\begin{theorem}
\label{thm:consistency_emp_selection_shrinkage}
Suppose \textbf{C1}--\textbf{C6} hold and $\frakF$ satisfies conditions (i)--(iv). Then,
\begin{itemize}
\item[(1)] $\sup_{\ell\in \calL}\sqrt{n}|\widehat{\Xi}(\ell,\widehat{h},\gamma_n) -\Xi(\ell,h,\gamma)| \stackrel{P}{\longrightarrow}0$ as $n\to\infty$.
\item[(2)] Let $\ell^*$ be any local maximizer of $\Xi(\ell,h,\gamma)$ in the interior of $\calL$. Assume there exists a neighborhood of $\ell^*$ such that for all feasible points $\ell\in \calL$ within the neighborhood, there exists a constant $\calK>0$ such that
\begin{equation}
\label{eq:converge_Xi}
\Xi(\ell,h,\gamma) -\Xi(\ell^*,h,\gamma) \leq -\calK \|\ell - \ell^* \|_2^2.
\end{equation}
Then, there exists a sequence ($\ell^*_{n}\colon n\in\mathbb{N}$) of local maximizers of $(\widehat{\Xi}(\ell,\widehat{h},\gamma_n)\colon n\in\mathbb{N})$ satisfying 
\begin{equation}
n^{1/4} \|\ell^*_{n} - \ell^* \|_2 = O_p(1) \qquad (n\to\infty). 
\label{eq:consistency_emp_selection_shrinkage_eq1}
\end{equation}
Further, recalling notation in Section \ref{sec:methodology}, under the null hypothesis, 
\begin{equation}
\frac{\sqrt{n}}{\widehat{\Delta}^{1/2}(f_{\ell^*_{n}} ,\gamma_n) } \{ \bM(f_{\ell^*_{n}})  - \widehat{\Omega}(f_{\ell^*_{n}},\gamma_n) I_q\} \Longrightarrow \bW.
\label{eq:consistency_emp_selection_shrinkage_eq2}  
\end{equation}
\item[(3)] Let $\ell^*$ be any local maximizer of~$\Xi(\ell,h,\gamma)$ on the boundary of $\calL$. Assume there exists a neighborhood of $\ell^*$ such that for all feasible points $\ell\in \calL$ within the neighborhood, there is a constant $\calK'>0$ satisfying 
\begin{equation}
\label{eq:converge_Xi_boundary} 
\Xi(\ell,h,\gamma) -\Xi(\ell^*,h,\gamma) \leq -\calK' \|\ell - \ell^* \|_2 .
\end{equation}
Then, (\ref{eq:consistency_emp_selection_shrinkage_eq1}) and (\ref{eq:consistency_emp_selection_shrinkage_eq2}) still hold. 
%Then, (\ref{eq:consistency_emp_selection_shrinkage_eq1}) and (\ref{eq:consistency_emp_selection_shrinkage_eq2}) still hold. 
\end{itemize}
\end{theorem}
The two conditions \eqref{eq:converge_Xi} and \eqref{eq:converge_Xi_boundary} ensure that the parameter $\ell^*$ is locally identifiable in a neighborhood of $\ell^*$. In general, the two conditions depend on the structure of $L^{\Sigma}$.

\subsection{Combination of prior models}
\label{subsec:combination_priors}

An extensive simulation analysis revealed that there is considerable variation in the shape of the power functions and the values of $\tilde{t} = (t_0,t_1,t_2)$, especially when the condition number of $\Sigma_p$ is relatively large. In this subsection, we consider a convenient collection of priors that are representative of certain structural scenarios. A composite test, called $\widehat{T}_{\max}$, is defined as the maximum of the standardized statistics $\widehat{T}(f_{\ell_i^*})$ where $\ell_i^*$ is obtained from Algorithm \ref{algo:empricial_selection_shrinkage} %the data-driven selection of parameters 
under prior $\tilde{t}_i$, $i=1,\dots, m$. %, using Algorithm \ref{algo:empricial_selection_shrinkage}. 
The following strategy is applicable to LR, LH and BNP. We therefore continue to use $\widehat{T}(f)$ to denote the general test statistic. In summary, we propose to test the hypothesis by rejecting for large values of the statistic
\[
\widehat{T}_{\max} = \max_{\tilde{t}\in \widetilde{\Pi}} \widehat{T}(f_{\ell_i^*} ), 
\]
where $\widetilde{\Pi} = \{\tilde{t}_1,\dots,\tilde{t}_m\}$, $m\geq 1$, is a pre-specified finite class of weights. A simple but effective choice of $\widetilde{\Pi}$ consists of the three canonical weights $\tilde{t}_1 = (1,0,0)$, $\tilde{t}_2 = (0,1,0)$, $\tilde{t}_3 = (0,0,1)$.

\begin{theorem}
\label{thm:asymp_normality_Tmax}
Suppose \textbf{C1}--\textbf{C6} hold and $\frakF$ satisfies condition (i)--(iv). For each $i= 1,\dots, m$, {assume that $\ell_{in}^*$ is a sequence of local maximizers of the empirical power function $\widehat{\Xi}(\ell,\widehat{h},\gamma_n)$ under prior model with weight $\tilde{t}_i$ such that 
\[n^{1/4} \|\ell_{in}^* - \ell_{i}^*\|_2  = O_p(1).\] (See (\ref{eq:consistency_emp_selection_shrinkage_eq1})).} Then, under the null hypothesis $H_0\colon BC= 0$,
\[ 
\big(\widehat{T}(f_{\ell_{1n}^*}),\dots,\widehat{T}(f_{\ell_{mn}^*}) \big) 
\Longrightarrow \mathcal{N}\big(0, \Delta_*\big),
\]
where $\Delta_*$ is an $m\times m$ matrix with diagonal entries $1$ and $(i,j)$-th off-diagonal entry \[
\Delta^{-1/2} (f_{\ell^*_{i}},\gamma) \Delta(f_{\ell^*_{i}},f_{\ell^*_{j}},\gamma) \Delta^{-1/2} (f_{\ell^*_{j}},\gamma).
\] 
\end{theorem}
Theorem \ref{thm:asymp_normality_Tmax} shows that $\widehat{T}_{\max}$ has a non-degenerate limiting distribution under $H_0$. It is worth mentioning that LR, LH and BNP share the covariance matrix $\Delta_*$. Theorem \ref{thm:asymp_normality_Tmax} can be used to determine the cut-off values of the test by deriving analytical formulas for the quantiles of the limiting distribution. Aiming to avoid complex calculations, a parametric bootstrap procedure is applied to approximate the cut-off values. Specifically, $\Delta_*$ is first estimated by $\widehat{\Delta}_*$, and then bootstrap replicates are generated by simulating from $\mathcal{N}(0, \widehat{\Delta}^*)$, thereby providing an approximation of the null distribution of $\widehat{T}_{\max}$. %A natural candidate for the covariance estimator is by 
Replacing $\Delta(f_{\ell^*_{i}},f_{\ell^*_{j}},\gamma)$ with $\widehat{\Delta}(f_{\ell^*_{i}},f_{\ell^*_{j}},\gamma_n)$ yields the natural estimator.

\begin{remark}
\label{remark:non_negative_Delta*}
Observe that $\widehat{\Delta}_*$ defined above may not be nonnegative definite even though it is symmetric. If such a case occurs, the resulting estimator can be projected onto its closest non-negative definite matrix simply by setting the negative eigenvalues to zero. This covariance matrix estimator is denoted by $\widehat{\Delta}_*^+$ and it is used for generating the bootstraps samples.
\end{remark}

\section{Ridge and higher-order regularizers}
\label{sec:ridge_regu_higher_order}
\subsection{Ridge regularization}
\label{subsec:ridge_regu}

One of the most commonly used shrinkage procedures in statistics is ridge regularization, corresponding to choosing $f_{\ell}(x) = 1/(x - \ell)$, $\ell <0$, so that $f_{\ell}(\hSigma) = (\hSigma - \ell I_p)^{-1}$. It is an effective way to shift $\hSigma$ away from singularity by adding a ridge term $-\ell I_p$. 
In this subsection, we apply the results of Sections \ref{sec:methodology} and \ref{sec:locally_most_powerful_test} using the ridge-shrinkage family
\[\frakF_{ridge} \coloneqq \{f_{\ell}(x) = (x -\ell)^{-1} ,\  \ell \in [\underline{\ell},\ \overline{\ell}]\}, \quad  -\infty<\underline{\ell}<\overline{\ell}<0.\] 
In the literature, ridge-regularization was applied to high-dimensional one- and two-sample mean tests in \citet{chen2011regularized} and \citet{LiAuePaulPengWang}. Hence, this subsection is a generalization of their methods to general linear hypotheses.

From the aspect of population covariance estimation, ridge-regularization can be viewed as an order-one estimation where $\Sigma_p$ is estimated by a weighted average of $\hSigma$ and $I_p$, namely $\alpha_0 I_p + \alpha_1 \hSigma $. The estimator is equivalent to ridge-regularization with  $\ell = {-\alpha_0/\alpha_1}$ for testing purposes. Within a restricted region of $(\alpha_1,\alpha_2)$, the large eigenvalues of $\hSigma$ are shrunk down and the small ones are lifted upward.  It is a desired property since in high-dimensional settings, large sample eigenvalues are systematically biased upward and small sample eigenvalues downwards. 

%The rest of the section is organized as following. In Section \ref{subsec:stat_formula}, we give the explicit formulas of test statistics and asymptotic local powers with ridge-shrinkage. In Section \ref{subsec:selection_local_alternatives}, we talk about the selection of local alternatives. 

An important advantage of ridge regularization is that the test procedure is computationally efficient due to the fact that $\Omega(f_{\ell},\gamma)$ and $\Delta(f_{\ell},\gamma)$ admit closed forms as shown in Lemma \ref{lemma:special_form_ARHT}. These quantities can be estimated by $\widehat{\Omega}_{\ell}(\gamma_n)=\widehat\Theta(\ell,\gamma_n)-1$ and $\widehat{\Delta}_{\ell}(\gamma_n)=2\widehat\delta(\ell,\ell,\gamma_n)$, respectively. %are consistent estimators of $\Omega(f_{\ell},\gamma)$ and $\Delta(f_{\ell},\gamma)$. 
A closed-form estimator $\widehat{\Xi}_\ell(\widehat{h},\gamma_n)$ is then also available for $\Xi({\ell}, h, \gamma)$. This leads to the following algorithm.
%For readers' convenience, we present the whole test procedure in the following algorithm, where we use $\widehat{\Omega}_{\ell}(\gamma_n)$, $\widehat{\Delta}_{\ell}(\gamma_n)$ and $\widehat{\Xi}_\ell(\widehat{h},\gamma_n)$ to denote those estimators of $\Omega(f_{\ell},\gamma)$, $\Delta(f_{\ell},\gamma)$ and $\Xi({\ell}, h, \gamma)$ when $f_{\ell} = (x-\ell)^{-1}$. 

\begin{algo}[Ridge-regularized test procedure] 
\label{algo:ridge_tests}
{~} \vspace{-.1cm}
\begin{itemize}\itemsep-.1ex
\item[1.] Specify prior weights $\tilde{t} = (t_0,t_1,t_2)$;
\item[2.] With  $m_{n,p}(\ell) = p^{-1}\tr(\hSigma -\ell I_p)^{-1}$, compute, for any $\ell\in [\underline{\ell}, \overline{\ell}]$,
\begin{align*}
\widehat{\Theta}({\ell}, \gamma_n) 
&= \{1-\gamma_n -\gamma_n\ell m_{n,p}(\ell)\}^{-1},\\
\widehat{\Omega}_{\ell}(\gamma_n) 
&= \widehat{\Theta}(\ell,\gamma_n) -1 ,\\
\widehat{\Delta}_\ell(\gamma_n)
&= 2\gamma_n\{1+ \ell m_{n,p}(\ell)\} \widehat{\Theta}^3(\ell,\gamma_n) + 2\gamma_n\ell \{ m_{n,p}(\ell) + \ell m'_{n,p}(\ell)\}\widehat{\Theta}^4(\ell, \gamma_n);
\end{align*}
\item[3.] For any $\ell\in [\underline{\ell},~ \overline{\ell}]$, compute $\widehat{h}(\ell,\gamma_n) = \sum_{j=0}^2 t_j \widehat{\rho}_j(\ell,\gamma_n)$ as defined in (\ref{eq:rho0_3}) and 
\[\widehat{\Xi}_{\ell}(\widehat{h},\gamma_n) = \frac{\widehat{h}(\ell,\gamma_n) }{\widehat\Delta^{1/2}_{\ell}(\gamma_n)};\]
\item[4.] Select $\ell^{*} = \mbox{arg}\max_{\ell\in [\underline{\ell}, ~\overline{\ell}]}\widehat{\Xi}_\ell(\widehat{h},\gamma_n)$;
\item[5.] Use one of the standardized statistics %(generically denoted here by $\widehat{T}$) to reject the null at asymptotic level $\alpha$, if $\widehat{T}> \xi_{\alpha}$.
\begin{align*}
\widehat{T}^{\rm LR}(\ell^{*})
& \coloneqq\frac{\sqrt{n}\{1+\widehat{\Omega}_{\ell^{*}}(\gamma_n)\} }{q^{1/2}\widehat{\Delta}^{1/2}_{\ell^{*}}(\gamma_n)} [T^{\rm LR}(\ell^{*}) - q \log\{1+\widehat{\Omega}_{\ell^{*}}(\gamma_n)\}],\\  
\widehat{T}^{\rm LH}(\ell^{*}) 
& \coloneqq\frac{\sqrt{n}}{q^{1/2}\widehat{\Delta}^{1/2}_{\ell^{*}}(\gamma_n)}[T^{\rm LH}(\ell^{*}) - q\widehat{\Omega}_{\ell^{*}}(\gamma_n)],\\  
\widehat{T}^{\rm BNP}(\ell^{*}) 
& \coloneqq\frac{\sqrt{n}\{1+\widehat{\Omega}_{\ell^{*}}(\gamma_n)\}^2}{q^{1/2}\widehat{\Delta}^{1/2}_{\ell^{*}}(\gamma_n)}\Big[T^{\rm BNP}(\ell^{*}) - \frac{q\widehat{\Omega}_{\ell^{*}}(\gamma_n)}{1+\widehat{\Omega}_{\ell^{*}}(\gamma_n)}\Big],
\end{align*}
where
\begin{align*}
T^{\rm LR}(\ell^{*}) = \sum\nolimits_{i=1}^q\log(1 + \lambda_i), \quad
T^{\rm LH}(\ell^{*}) =   \sum\nolimits_{i=1}^q \lambda_i, \quad
T^{\rm BNP}(\ell^{*}) = \sum\nolimits_{i=1}^q \frac{\lambda_i}{1+\lambda_i},
\end{align*}
and $\lambda_1,\ldots,\lambda_q$ are the eigenvalues of $n^{-1} Q_n^T \bY^T(\hSigma - \ell^{*} I_p)^{-1} \bY Q_n$. Reject the null at asymptotic level $\alpha$ if the test statistic value exceeds $\xi_\alpha$. % $\widehat{T}> \xi_{\alpha}$.
\end{itemize}
\end{algo}

Although in theory any negative $\ell^*$ is allowed in the test procedure, in practice, meaningful lower and upper bounds $\underline{\ell}$ and $\overline{\ell}$ are needed to ensure stability of the test statistics when $p \approx n$ or $p>n$ and also to carry out the search for optimal $\ell$ at a low computational cost. In our simulation settings we use $\overline{\ell} = - p^{-1} \tr(\hSigma)/100$ and $\underline{\ell} = - 20\lambda_{\max}(\hSigma)$, which generally lead to quite robust performance.

\begin{table}[htbp]
\setlength{\belowcaptionskip}{-10pt}
\begin{center}
\scriptsize
\resizebox{\textwidth}{!}{
%\begin{tabular}{cccccccccccccc}
\begin{tabular}{cc@{\quad}ccc@{\quad}ccc@{\quad}ccc@{\quad}ccc}
\hline
 \multicolumn{2}{c}{\multirow{2}{*}{}}%{Size $\times$ 100\%}}                                   
 & \multicolumn{6}{c}{$\Sigma  =I_p$}                  & \multicolumn{6}{c}{$\Sigma = \Sigma_{den}$}        \\ %\hline %\cline{3-14}
                           &                       & \multicolumn{3}{c}{$k =3$} & \multicolumn{3}{c}{$k=5$} & \multicolumn{3}{c}{$k =3$} & \multicolumn{3}{c}{$k=5$} \\ %\cline{1-14} 
                            \multicolumn{2}{c}{$n=300$,\ $p\ =$} & 150    & 600    & 3000   & 150    & 600    & 3000  & 150    & 600    & 3000   & 150     & 600   & 3000  \\ \cline{1-14}

\multirow{3}{*}{LR$_{\rm ridge}$}  & $\tilde{t}_1 $    & 5.4   & 5.2   & 5.1   & 5.2   & 5.1   & 5.1  & 4.9   & 4.4   & 4.7   & 4.4    & 3.3  & 4.2  \\  
                           & $\tilde{t}_2 $    & 5.4   & 5.2   & 5.1   & 5.2   & 5.1   & 5.1  & 4.9   & 5.2   & 4.9   & 4.4    & 4.9  & 4.7  \\ 
                           & $\tilde{t}_3 $    & 5.3   & 5.2   & 5.1   & 5.2   & 5.1   & 5.1  & 5.8   & 5.9   & 5.1   & 5.3    & 5.2  & 4.9  \\[.1cm]
\multirow{3}{*}{LH$_{\rm ridge}$}  & $\tilde{t}_1 $    & 5.4   & 5.2   & 5.1   & 5.3   & 5.1   & 5.2  & 6.2   & 7.2   & 5.7   & 6.2    & 7.7  & 6.0  \\ 
                           & $\tilde{t}_2 $    & 5.4   & 5.2   & 5.1   & 5.3   & 5.1   & 5.2  & 6.2   & 5.9  & 5.2   & 6.2    & 5.9  & 5.1  \\
                           & $\tilde{t}_3 $    & 5.3   & 5.2   & 5.1   & 5.3   & 5.1   & 5.2  & 5.8   & 5.9   & 5.2   & 5.4    & 5.2  & 5.0  \\[.1cm]
\multirow{3}{*}{BNP$_{\rm ridge}$} & $\tilde{t}_1 $    & 5.3   & 5.2   & 5.0   & 5.2   & 5.0   & 5.0  & 4.0   & 2.5   & 3.7   & 2.9   & 1.3  & 3.1  \\ 
                           & $\tilde{t}_2 $    & 5.4   & 5.2   & 5.0   & 5.2   & 5.0   & 5.0  & 4.0   & 4.7   & 4.6   & 2.9    & 3.9  & 4.4  \\ 
                           & $\tilde{t}_3 $     & 5.3   & 5.2   & 5.0   & 5.2   & 5.0   & 5.0  & 5.8   & 5.8   & 5.0   & 5.3   & 5.1  & 4.7  \\ \hline
\multirow{3}{*}{LR$_{\rm high}$}  & $\tilde{t}_1 $   &6.5  &6.3   &5.3     &6.5    & 5.3  & 5.5   & 6.0   & 5.8   & 5.1   & 6.5   &5.9 &   4.5  \\ 
                                            %& Up 6.2   &6.0   &5.4   & 5.7   & 5.8   & 5.7  & 6.2   & 7.2   & 5.7   & 6.2    & 7.7  & 5.9  \\
                           & $\tilde{t}_2 $    &6.5  & 6.3  & 5.3  & 6.5 & 5.3 & 5.5 & 8.3  & 6.8  & 5.5  & 8.4  & 7.2  & 5.2\\
                                          %&  Up 6.2   & 6.0   & 5.4   & 5.7   & 5.8   & 5.7  & 6.2   & 7.2   & 6.0  & 6.2    & 7.3  & 6.1  \\ 
                           & $\tilde{t}_3 $    &6.6 &6.3  &5.3  &6.6  &5.3  &5.5  &6.7  &6.7  &5.5  &6.4  &7.1  &5.2 \\[.1cm]
                                         %&  Up 6.2   & 6.0   & 5.4   & 5.7   & 5.8   & 5.7  & 6.6   & 7.1   & 5.6  & 6.5    & 6.9  & 5.7  \\ \hline
\multirow{3}{*}{LH$_{\rm high}$}  & $\tilde{t}_1 $  &6.7 &6.4 &5.4  &6.8  &5.5  &5.7  &6.1  &5.9  &5.7  &6.7  &6.2  &5.5 \\
                                          % Up & 6.3   & 6.1   & 5.4   & 5.9   & 5.9   & 5.7  & 6.2   & 7.2   & 5.7   & 6.2    & 7.7  & 6.0  \\ 
                           & $\tilde{t}_2 $     &6.7  &6.4  &5.4 &6.8 &5.4  &5.7  &8.3  &6.8  &5.6  &8.5  &7.3  &5.5\\
                           %& Up 6.3   & 6.1   & 5.4   & 5.9   & 5.9   & 5.7  & 6.2   & 7.3   & 6.0   & 6.2    & 7.3  & 6.1  \\
                           & $\tilde{t}_3 $    &6.7   &6.4 &5.4  &6.8 &5.4  &5.7  &6.7  &6.7  &5.6  &6.5  &7.2   &5.5 \\[.1cm]
                           %& Up 6.3   & 6.1   & 5.4   & 5.9   & 5.9   & 5.7  & 6.6   & 7.1   & 5.6   & 6.5    & 6.9  & 5.7  \\ \hline
\multirow{3}{*}{BNP$_{\rm high}$} & $\tilde{t}_1 $  &6.2  &6.3  &5.2  &6.1  &5.3  &5.2  &5.9  &5.7  &4.6  &6.4  &5.5  &3.7 \\
                            %&Up 6.0   & 5.9   & 5.4   & 5.4   & 5.7   & 5.7  & 6.2   & 7.2   & 5.6   & 6.2    & 7.7  & 5.9  \\  
                           & $\tilde{t}_2 $    &6.3   &6.3 &5.2  &6.1  &5.2  &5.2  &8.3  &6.7  &5.3  &8.3  &7.0  &4.9 \\
                           %& Up 6.0   & 5.9   & 5.4   & 5.4   & 5.7   & 5.6  & 6.2   & 7.2   & 5.9   & 6.2    & 7.2  & 6.1  \\
                           & $\tilde{t}_3 $    &6.3   &6.3  &5.1  &6.1  &5.2  &5.2  &6.6  &6.6  &5.3  &6.4  &6.9  &4.9 \\ \hline
\multicolumn{2}{l}{~LR$_{\rm comp}$}               &5.1   &5.1  &5.0  &5.4  &5.3  &5.0  &6.0  &5.1  &5.5  &5.6  &5.0  &5.1 \\[.1cm]
\multicolumn{2}{l}{~LH$_{\rm comp}$}               &5.1   &5.1  &5.1  &5.5  &5.3  &5.1  &6.7  &5.8  &5.9  &6.9  &6.2  &5.7 \\[.1cm]
\multicolumn{2}{l}{~BNP$_{\rm comp}$}              &5.1   &5.0  &5.0  &5.4  &5.2  &5.0  &5.4  &4.5  &5.1  &4.7  &4.4  &4.6\\ \hline
                           %& Up 6.0   & 5.9   & 5.4   & 5.4   & 5.7   & 5.7  & 6.6   & 7.0   & 5.6   & 6.5    & 6.9  & 5.7  \\ \hline
\multicolumn{2}{l}{~ZGZ}                      &5.6 &5.7 &5.2 &5.6 &4.8 &5.2  &5.9  &5.5  &5.4 &5.4 &5.4 &5.3\\ %\hline
%&5.3   & 5.2   & 5.0   & 5.1   & 5.0   & 5.1  & 5.7   & 5.8  & 5.2   & 5.2    & 5.0  & 4.9  \\ \hline
\multicolumn{2}{l}{~CX (Oracle)}               &5.6   & 6.3   &7.0   &7.3    &6.9   & 8.6   &5.8  &5.9  &6.8 &6.0  &7.2  &9.0 \\ \hline
%&5.3   & 5.9   & 6.6   & 6.9   & 7.2   & 8.5  & 0.4   & 0.2   & 0.1   & 0.5    & 0.3  & 0.1  \\ \hline
\end{tabular}}
\caption{\footnotesize Empirical sizes at level $5\%$. $\Sigma = \mbox{ID}$ and $\Sigma_{den}$; $\tilde{t}_1 = (1,0,0),$ $\tilde{t}_2 = (0,1,0)$, $\tilde{t}_3= (0,0,1)$.}
\label{table:emp_size_ID_den}
\end{center}
\end{table}

The composite test procedure with ridge-regularization is summarized below.     %in Algoritm \ref{algo:composite_ridge_test}. 
\begin{algo}[Composite ridge-regularized test procedure] 
\label{algo:composite_ridge_test}
{~} \vspace{-.1cm}
\begin{itemize}\itemsep-.1ex
\item[1.] Select prior weights $\widetilde{\Pi} = (\tilde{t}_1,\dots,\tilde{t}_m) $.  The canonical choice is  $((1,0,0), (0,1,0),\break (0,0,1))$;
\item[2.] For each $\tilde{t}_j$ in $\widetilde{\Pi}$, run Algorithm \ref{algo:ridge_tests}, get the standardized test statistic $\widehat{T}(\ell^*_j)$ and compute $\widehat{T}_{\max} = \max_{1\leq j \leq m} \widehat{T}(\ell^*_j)$;
\item[3.] With the selected tuning parameters $(\ell^*_1,\ell^*_2,\ell^*_3)$ compute the matrix $\widehat{\Delta}_*$ whose diagonal elements are equal to one and whose $(i,j)$-th entry for $i\neq j$ is
\[\widehat{\Delta}^{-1/2}_{\ell^*_i}(\gamma_n)\widehat{\Delta}_{\ell^*_i,\ell^*_j}(\gamma_n)\widehat{\Delta}^{-1/2}_{\ell^*_j}(\gamma_n),\] where $\widehat{\Delta}_{\ell^*_i}(\gamma_n)$ is defined in Step 2 of Algorithm \ref{algo:ridge_tests} and 
\begin{align*}
\widehat{\Delta}_{\ell^*_i,\ell^*_j}(\gamma_n) = 2\widehat{\Theta}({\ell^*_i}, \gamma_n) \widehat{\Theta}({\ell^*_j}, \gamma_n) \Big[\frac{\ell^*_i \widehat{\Theta}({\ell^*_i}, \gamma_n)- \ell^*_j\widehat{\Theta}({\ell^*_j}, \gamma_n) }{\ell^*_i -\ell^*_j }-1\Big];
\end{align*}
%The diagonal element of $\widehat{\Delta}_*$ is 1.
\item[4.] Project $\widehat{\Delta}_*$ to its closest non-negative definite matrix $\widehat{\Delta}^+_*$  by setting the negative eigenvalues to zero. Generate $\varepsilon_1,\dots, \varepsilon_{G}$ with $\varepsilon_{b} = \max_{1\leq i\leq m} Z_i^{(b)}$ with $Z^{(b)} = [Z_i^{(b)}]_{i=1}^m \sim \mathcal{N}(0,\widehat{\Delta}^+_*)$. 
\item[5.] Compute the $p$-value as $G^{-1} \sum_{b=1}^G \mathbbm{1}\{\varepsilon_b >  \widehat{T}_{\max}\}$.
\end{itemize}
\end{algo}

\subsection{Extension to higher-order regularizers}
\label{subsec:higher_order_extension}

Through an extensive simulation study in a MANOVA setting, it is shown in Section \ref{sec:simulation} that the ridge-regularized tests compare favorably against a host of existing test procedures. This is consistent with the findings in \citet{LiAuePaulPengWang} in the two-sample mean test framework. %Especially, it shows advantages against \citet{zhou2017high,bai1996effect,chen2010two}. The latter can be viewed as order-zero procedures where they estimate $\Sigma_p$ by $\alpha_0 I_p$. 
Ridge-shrinkage rescales $\widehat{\bH}_p$ by $(\hSigma- \ell I_p)^{-1}$ instead of $\hSigma^{-1}$. Broader classes of scaling matrices have been studied extensively \citep[see][for an overview]{ledoit2012nonlinear}. They can be set up in the form $f(\hSigma)$. When $f(\cdot)$ is analytic, such scaling falls within the class of the proposed tests.

The flexibility provided by a larger class of scaling matrices can be useful to design test procedures for detecting a specific kind of alternative. The choice of the test procedure may for example be guided by questions such as \emph{Which $f$ leads to the best asymptotic power under
a specific sequence of local alternatives, if $H_0$ is rejected based on large eigenvalues of $\bM(f)$?}
While a full characterization of this question is beyond the scope of this paper, a partial answer may be provided by restricting to functions $f$ in the higher-order class
\[
\frakF_{{\rm high}} =  \Big\{ f_{\ell}(x) = \Big[\sum\nolimits_{j=0}^\kappa l_j x^j\Big]^{-1}\colon \ell = (l_0, \dots, l_\kappa)^T \in \calG \Big\}, 
\]
where $\calG$ is such that $f_{\ell}$ is uniformly bounded and monotonically decreasing on $\calX$, for any $\ell\in \calG$. These higher-order shrinkage functions are weighted averages of ridge-type shrinkage functions. To see this, suppose the polynomial $\sum_{j=0}^\kappa l_j x^j$ has roots $r_1,\dots, r_{\kappa_0}\in\mathbb{C}\setminus\calX$ with multiplicity $s_1,\dots, s_{\kappa_0}\in\mathbb{N}^+$. Via basic algebra, $f_{\ell}$ can be expressed as  
\begin{equation}
\label{eq:weight_average_repexression_f}
f_{\ell}(x) = \Big[\sum\nolimits_{j=0}^\kappa l_j x^j\Big]^{-1} = \sum\nolimits_{j=1}^{\kappa_0}\sum\nolimits_{i=1}^{s_{\kappa_0}} w_{ji} (x -r_j)^{-i},
\end{equation}
with some weights $w_{ji}\in\mathbb{C}$. If all roots are simple, $f_{\ell}$ is a weighted average of ridge-regularization with  $\kappa$ different parameters. Heuristically, it is expected that a higher order $f_{\ell}$ yields tests more robust against unfavorable selection of ridge shrinkage parameter.  

The design of $\calG$ is not easy when $\kappa$ is large. Here, we select $\kappa = 3$, which is the minimum degree that allows $f^{-1}_{\ell}$ to be both locally convex and concave. In this case, the complexity of selecting the optimal regularizer is significantly higher than for ridge-regularization. Due to space limitations, we move the design of $\calG$ and the test procedure when $\kappa=3$ to Section \ref{sec:add_material_higher_order} of the Supplementary Material.

\section{Simulations}
\label{sec:simulation}

In this section, the proposed tests are compared by means of a simulation study to two representative existing methods in the literature, \citet{zhou2017high} (ZGZ) and \citet{cai2014high} (CX). We focus on one-way MANOVA, a set-up for which both competing methods are applicable. It is worth mentioning that CX requires a good estimator of the precision matrix $\Sigma_p^{-1}$, that is typically unavailable when both $\Sigma_p$ and $\Sigma_p^{-1}$ are dense. In the simulations, the true $\Sigma_p^{-1}$ is utilized for CX, thus making it an oracle procedure.  
In the following, LR$_{\rm ridge}$, LH$_{\rm ridge}$, and BNP$_{\rm ridge}$ denote the ridge-regularized tests presented in Algorithm \ref{algo:ridge_tests}. LR$_{\rm high}$, LH$_{\rm high}$, and BNP$_{\rm high}$ denote the tests with higher-order shrinkage introduced in Section \ref{subsec:higher_order_extension} %and Section \ref{sec:add_material_higher_order} in the Supplementary Material 
with $\kappa =3$. LR$_{\rm comp}$, LH$_{\rm comp}$ and BNP$_{\rm comp}$ denote the composite ridge-regularized tests of Algorithm \ref{algo:composite_ridge_test} with the canonical choice of $\widetilde{\Pi}= ((1,0,0), (0,1,0), (0,0,1))$.

\subsection{Settings}
The observation matrix $\bY$ was generated as in (\ref{eq:linear_model}) with normally distributed $\bZ$. Specifically, we selected $k =3$ or $5$, and $N =300$. For $k=3$, the three groups had $75, 90$ and $135$ observations, respectively. For $k=5$, the design was balanced with each group containing $60$ observations. The dimension $p$ was $150, 600, 3000$, so that $\gamma_n = p/n \approx 0.5, 2$ and $10$. The columns of $B$ were the $k$ group mean vectors. Accordingly, the columns of $X$ were the group index indicators of observation subjects. We selected $C$ to be the successive contrast matrix of order $q = k-1$. This is a standard one-way MANOVA setting.

% Figure 1 
\begin{figure}[htbp]
\setlength{\belowcaptionskip}{-5pt}
\includegraphics[width = \linewidth,height = 0.4\linewidth]{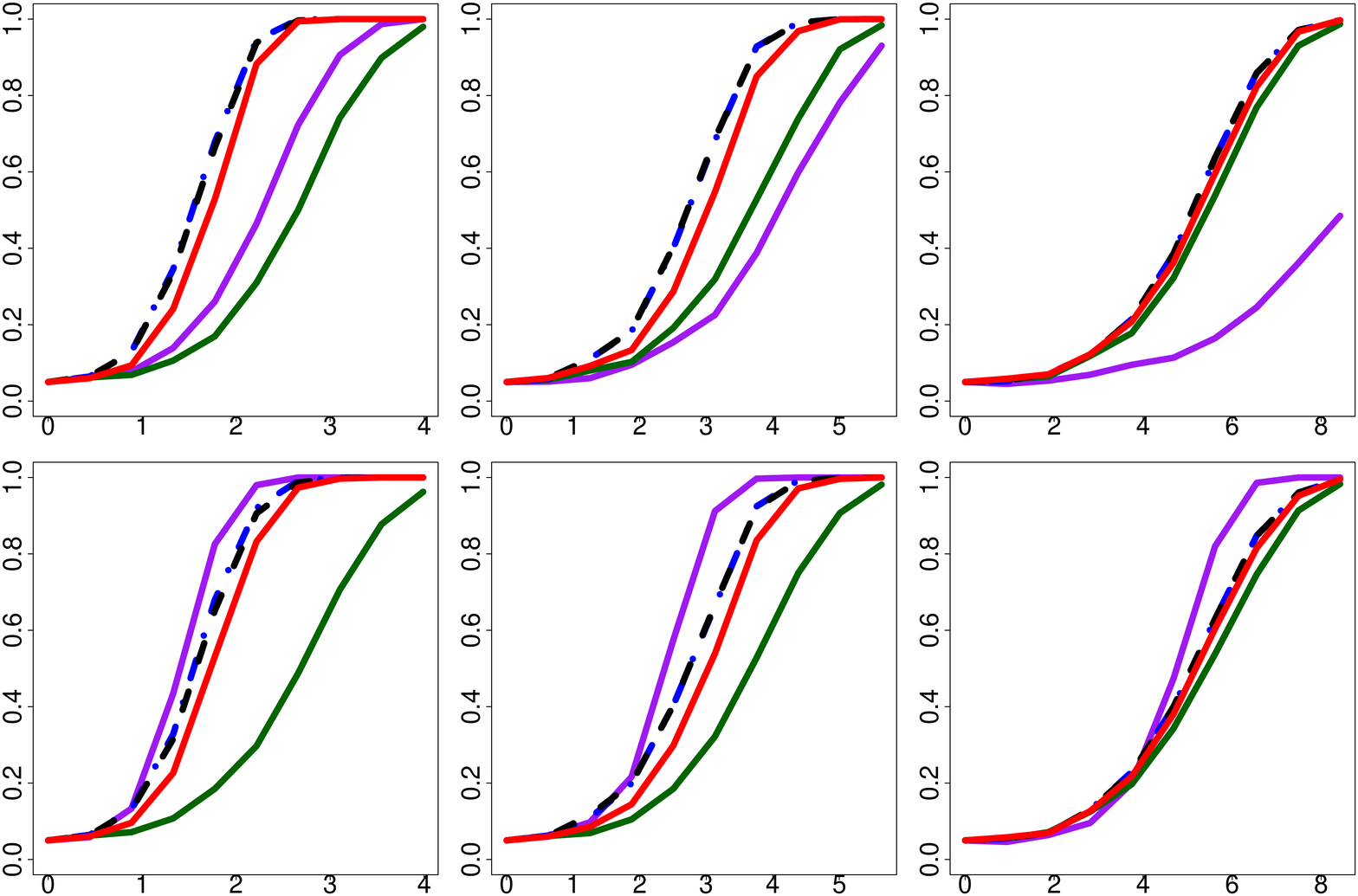}
\caption{\footnotesize Size-adjusted power with $\Sigma=\Sigma_{den}$,~$k=5$. Rows (top to bottom): $B =\mbox{Dense and Sparse} $; Columns (left to right): $p = 150, 600, 3000$. BNP$_{\rm comp}$ (red, solid);  ZGZ (green, solid); oracle CX (purple, solid); BNP$_{\rm ridge}$ (black, dashed) and BNP$_{\rm high}$ (blue, dotted-dashed) with $\tilde{t} = (1,0,0)$.}
\label{fig:SA_Sigma_dense_k5_t100_BNP}
\end{figure}

Under the null, $B$ is the zero matrix. Under the alternative, for each setting of the parameters and each replicate, $B$ is generated using one of the following models. 
\begin{itemize}
\item[(i)] \emph{Dense alternative}: The entries of $B$ are i.i.d.\ $\mathcal{N}(0,c^2)$ with $c = O(n^{-1/4}p^{-1/2})$ used to tune signal strength to a non-trivial level. 
\item[(ii)] \emph{Sparse altenative}: $B = c \bR \bcalV$ with $c = O(n^{-1/4}p^{-1/2})$, where $\bR$ is a diagonal $p\times p$ matrix with 10\% randomly and uniformly selected diagonal entries being $\sqrt{10}$ and the remaining 90\% being equal to 0, and $\bcalV$ is a $p\times p$ matrix with i.i.d.\ standard normal entries.  %While the first column of $B$ is always $0$, each other column has $10\%$ randomly selected non-zero entries. The nonzero entries in Column $j$ have a uniform distribution on the support ${a}_{\cdot j}^{-1}\{a_{1,j},\dots, a_{(0.1p)j}\}$ where $a_{ij}$ is uniformly sampled from $\{1,\dots,p\}$ with replacement and ${a}_{\cdot j} = (p^{-1}{\sum_i a_{i,j}^2})^{1/2} $. Afterwards, $B$ is scaled by $c$ with $c = O(n^{-1/4}p^{-1/2})$. It's a setting borrowed from \citet{cai2014high}.   
\end{itemize}
The following four models for the covariance matrix $\Sigma =\Sigma_p$ were considered. All models were further scaled so that $\tr(\Sigma_p) = p$.
\begin{itemize}
\item[(i)]  \emph{Identity matrix} (ID): $\Sigma = I_p$.
\item[(ii)] \emph{Dense case} $\Sigma_{den}$: Here $\Sigma= P \Sigma_{(1)} P^T$ with a unitary matrix $P$ randomly generated from the Haar measure and resampled for each different setting, and a diagonal matrix $\Sigma_{(1)}$ whose eigenvalues are given by $\lambda_j = (0.1+j)^6  + 0.05  p^6$, $j = 1,\dots,p$. The eigenvalues of $\Sigma$ decay slowly, so that no dominating leading eigenvalue exists.
\item[(iii)] \emph{Toeplitz case} $\Sigma_{toep}$: Here $\Sigma$ is a Teoplitz matrix with the $(i,j)$-th element equal to $0.5^{|i-j|}$. It is a setting where $\Sigma^{-1}$ is sparse but $\Sigma$ is dense.
\item[(iv)] \emph{Discrete case} $\Sigma_{dis}$: Here  $\Sigma= P \Sigma_{(2)} P^T$ with $P$ generated in the same way as in (ii), and $\Sigma_{(2)}$ is a diagonal matrix with 40\% eigenvalues 1, 40\% eigenvalues 3 and 20\% eigenvalues 10. 
\end{itemize} 

\begin{table}[htbp]
\setlength{\belowcaptionskip}{-10pt}
\begin{center}
\scriptsize
\resizebox{\textwidth}{!}{
%\begin{tabular}{cccccccccccccc}
\begin{tabular}{cc@{\quad}ccc@{\quad}ccc@{\quad}ccc@{\quad}ccc}
\hline
 \multicolumn{2}{c}{\multirow{2}{*}{}}%{Size $\times$ 100\%}}                                   
 & \multicolumn{6}{c}{$\Sigma  =\Sigma_{dis}$}                  & \multicolumn{6}{c}{$\Sigma = \Sigma_{toep}$}        \\ %\hline %\cline{3-14}
                           &                       & \multicolumn{3}{c}{$k =3$} & \multicolumn{3}{c}{$k=5$} & \multicolumn{3}{c}{$k =3$} & \multicolumn{3}{c}{$k=5$} \\ %\cline{1-14} 
                            \multicolumn{2}{c}{$n=300$,\ $p\ =$} & 150    & 600    & 3000   & 150    & 600    & 3000  & 150    & 600    & 3000   & 150     & 600   & 3000  \\ \cline{1-14}

\multirow{3}{*}{LR$_{\rm ridge}$}& $\tilde{t}_1 $   &  4.8 & 5.0 & 4.6 & 4.7 & 4.5 & 5.0 & 5.4 & 4.4 & 4.8 & 4.5 & 4.6 & 4.6 \\
& $\tilde{t}_2 $  &5.1 & 5.2 & 4.9 & 5.2 & 4.6 & 5.1 & 5.4 & 4.9 & 4.9 & 4.9 & 4.8 & 5.0 \\
& $\tilde{t}_3 $  &5.6 & 5.5 & 5.1 & 5.7 & 5.3 & 5.3 & 5.8 & 5.2 & 5.0 & 5.7 & 5.4 & 5.1 \\[.1cm]
\multirow{3}{*}{LH$_{\rm ridge}$}& $\tilde{t}_1 $ &5.8 & 6.0 & 5.2 & 6.6 & 6.3 & 5.6 & 6.4 & 5.3 & 5.2 & 6.2 & 6.3 & 5.3 \\
& $\tilde{t}_2 $  & 5.7 & 5.7 & 5.1 & 6.3 & 5.6 & 5.5 & 5.9 & 5.3 & 5.0 & 5.8 & 5.6 & 5.3 \\
& $\tilde{t}_3 $  &5.6 & 5.5 & 5.2 & 5.8 & 5.3 & 5.4 & 5.8 & 5.3 & 5.1 & 5.7 & 5.4 & 5.2 \\[.1cm]
\multirow{3}{*}{BNP$_{\rm ridge}$}& $\tilde{t}_1 $ & 3.9 & 4.1 & 4.3 & 3.1 & 3.1 & 4.1 & 4.4 & 3.7 & 4.4 & 3.2 & 3.4 & 3.9 \\
& $\tilde{t}_2 $  &4.6 & 4.8 & 4.8 & 4.1 & 4.0 & 4.9 & 4.9 & 4.4 & 4.8 & 4.1 & 4.3 & 4.7 \\
& $\tilde{t}_3 $  &5.5 & 5.5 & 5.0 & 5.7 & 5.2 & 5.1 & 5.8 & 5.2 & 5.0 & 5.6 & 5.4 & 5.1 \\
\hline
\multirow{3}{*}{LR$_{\rm high}$}  & $\tilde{t}_1 $ &6.3 &6.4  &4.8  &5.9  &7.0  &5.5  &7.1  &7.0  &5.3  &7.5  &6.9  &5.2\\
                                            %& Up 5.8 & 6.0 & 5.6 & 6.6 & 6.2 & 6.0 & 6.4 & 5.3 & 5.7 & 6.2 & 6.3 & 5.8 \\
                             & $\tilde{t}_2 $  &7.9 &6.5  &4.8  &8.3  &7.1  &5.5  &7.6  &7.2  &5.3  &7.8  &7.0  &5.2\\
                                            %& Up 5.9 & 6.6 & 5.6 & 6.3 & 6.5 & 6.0 & 6.5 & 6.3 & 5.7 & 6.2 & 6.4 & 5.8 \\
                             & $\tilde{t}_3 $ & 6.1 &5.6  &4.8  &6.4  &6.1  &5.5  &6.7  &6.5  &5.3  &6.6  &6.4  &5.2\\[.1cm]
                                            %& Up 6.2 & 6.0 & 5.5 & 6.5 & 5.9 & 5.9 & 6.6 & 5.7 & 5.6 & 6.4 & 6.0 & 5.7 \\
\multirow{3}{*}{LH$_{\rm high}$}  & $\tilde{t}_1 $ & 6.6 &6.5  &5.0 &6.2 &7.2  &5.7  &7.2  &7.2  &5.5  &7.7  &7.0  &5.5\\
                                            %& Up 5.8 & 6.0 & 5.7 & 6.6 & 6.3 & 6.1 & 6.4 & 5.3 & 5.7 & 6.2 & 6.3 & 5.9 \\
                             & $\tilde{t}_2 $  &8.0  &6.6  &5.0  &8.5  &7.2  &5.7  &7.8  &7.2  &5.5 &8.0  &7.1 &5.5\\
                                            %& Up 5.9 & 6.6 & 5.7 & 6.4 & 6.6 & 6.1 & 6.6 & 6.3 & 5.7 & 6.4 & 6.5 & 5.9 \\
                            & $\tilde{t}_3 $   &6.2  &5.6  &5.0  &6.5  &6.2  &5.7  &6.7  &6.5  &5.5  &6.7  &6.5 &5.5\\[.1cm]
                                            %& Up 6.2 & 6.0 & 5.5 & 6.6 & 6.0 & 5.9 & 6.6 & 5.7 & 5.6 & 6.5 & 6.1 & 5.8 \\
\multirow{3}{*}{BNP$_{\rm high}$}  & $\tilde{t}_1 $ &6.1 &6.3  &4.7  &5.6  &6.8  &5.3  &7.1  &7.0  &5.2  &7.2  &6.8  &5.1\\
                                            %&5.8 & 5.9 & 5.6 & 6.6 & 6.2 & 6.0 & 6.4 & 5.3 & 5.7 & 6.2 & 6.2 & 5.8 \\
                              & $\tilde{t}_2 $  &7.9 &6.4  &4.7  &8.2  &7.0  &5.3  &7.5  &7.1  &5.2  &7.7  &7.0  &5.1\\
                                            %& 5.8 & 6.5 & 5.6 & 6.2 & 6.5 & 6.0 & 6.4 & 6.3 & 5.7 & 5.9 & 6.3 & 5.8 \\
                              &$\tilde{t}_3 $   &6.1 &5.5  &4.7  &6.4  &6.0  &5.3  &6.6  &6.4  &5.2  &6.5  &6.3  &5.1\\
                                            %&Up &6.1 & 5.9 & 5.5 & 6.3 & 5.9 & 5.9 & 6.5 & 5.6 & 5.6 & 6.3 & 6.0 & 5.7 \\
\hline

\multicolumn{2}{l}{~LR$_{\rm comp}$}  &6.2 &5.2 &5.0 &5.2 &5.3 &5.5 &5.9 &5.0 &5.1 &5.5 &4.9 &4.9\\[.1cm]
\multicolumn{2}{l}{~LH$_{\rm comp}$}  &7.0 &5.9 &5.3 &6.5 &6.4 &6.0 &6.6 &5.6 &5.3 &6.6 &5.7 &5.3\\[.1cm]
\multicolumn{2}{l}{~BNP$_{\rm comp}$} &5.5 &4.6 &4.8 &4.4 &4.6 &5.0 &5.4 &4.6 &4.9 &4.8 &4.4 &4.6\\ \hline
\multicolumn{2}{l}{~ZGZ}  &5.5  &4.7  &4.6  &5.7  &5.1  &5.3  &6.0  &5.5  &5.0  &5.9  &5.6  &5.0 \\[.1cm] 
%& 5.6 & 5.5 & 5.1 & 5.6 & 5.2 & 5.3 & 5.8 & 5.3 & 5.0 & 5.6 & 5.3 & 5.1 \\
\multicolumn{2}{l}{~CX (Oracle) }  &5.3 &5.9 &6.6 &6.8 &7.2  &8.6 &5.3 &6.2 &6.8 &6.8 &7.2 &8.4\\
%&5.7 & 6.3 & 6.9 & 6.5 & 7.5 & 8.5 & 5.4 & 5.9 & 6.8 & 6.6 & 7.3 & 8.5\\
\hline
\end{tabular}}
\caption{\footnotesize Empirical sizes at level $5\%$. $\Sigma = \Sigma_{dis}$ and $\Sigma_{toep}$; $\tilde{t}_1 = (1,0,0),$ $\tilde{t}_2 = (0,1,0)$, $\tilde{t}_3= (0,0,1)$.}
\label{table:emp_size_dis_toep}
\end{center}
\end{table}

All tests were conducted at significance level $\alpha = 0.05$. Empirical sizes for the various tests are shown in Tables \ref{table:emp_size_ID_den} and \ref{table:emp_size_dis_toep}.  Empirical power curves versus expected signal strength $n^{1/4}p^{1/2}c$ are reported in Figures \ref{fig:SA_Sigma_dense_k5_t100_BNP}--\ref{fig:SA_Sigma_toep_k3_t010_LR}. To better compare the power of each test, curves are displayed after size adjustment where the tests utilize the size-adjusted cut-off values based on the actual null distribution computed by simulations. Counterparts of Figures \ref{fig:SA_Sigma_dense_k5_t100_BNP}--\ref{fig:SA_Sigma_toep_k3_t010_LR} that utilize asymptotic (approximate) cut-off values are reported in Section \ref{sec:nominal_simulation} of the Supplementary Material.
 The difference between the two types is limited. LR, LH and BNP criteria behave similarly across simulation settings, as indicated by Theorem \ref{thm:probability_local_power_three_stat}. Therefore, only one of them is displayed in each figure for ease of visualization. More figures can be found in Section \ref{sec:additional_simulation} of the Supplementary Material. Note that, in some of the settings, several of the power curves nearly overlap, creating an occlusion effect. Then, power
curves corresponding to the composite tests are plotted as the top layer.

\subsection{Summary of simulation results}
\label{subsec:summary_simulation}
Tables \ref{table:emp_size_ID_den} and \ref{table:emp_size_dis_toep} show the empirical sizes of the proposed tests are mostly controlled under 7.5\%. The slight oversize is caused by the fact that $\bM(f)$ behaves like a quadratic form, therefore the finite sample distribution is skewed. LR and BNP tests are more conservative than LH tests because the former two calibrate the statistics by transforming eigenvalues of $\bM(f)$. Ridge-regularized tests are slightly more conservative under higher-order shrinkage. %A possible reason is that the more parameters in higher order shrinkage thicken the tail of the statistic with the data-driven shrinkage. 

% Figure 2 
\begin{figure}[htbp]
\setlength{\belowcaptionskip}{-10pt}
\includegraphics[width = \linewidth,height = 0.4\linewidth]{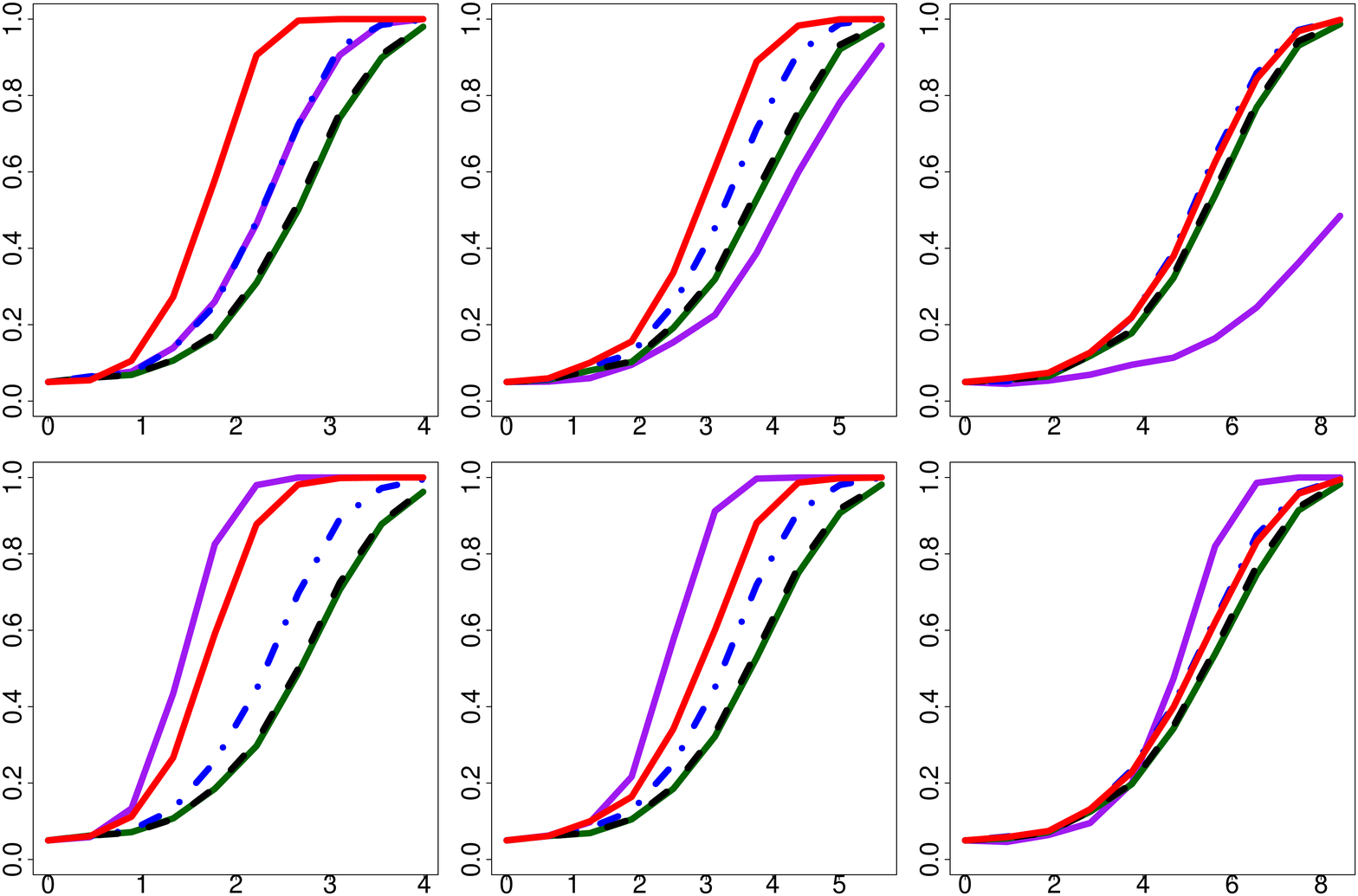}
\caption{\footnotesize Size-adjusted power with $\Sigma=\Sigma_{den}$,~$k=5$. Rows (top to bottom): $B =\mbox{Dense and Sparse} $; Columns (left to right): $p = 150, 600, 3000$. LH$_{\rm comp}$ (red, solid);  ZGZ (green, solid); oracle CX (purple, solid); LH$_{\rm ridge}$ (black, dashed) and LH$_{\rm high}$ (blue, dotted-dashed) with $\tilde{t} = (0,0,1)$.}
\label{fig:SA_Sigma_dense_k5_t001_LH}
\end{figure}

Note that in both simulation settings, $B$ consists of independent entries. Therefore, $\tilde{t}_1 = (1,0,0)$ is considered as a correctly specified prior, while $\tilde{t}_2 = (0,1,0)$ and $\tilde{t}_3 = (0,0,1)$ are considered as moderately and severely misspecified, respectively. {The composite tests combine $\tilde{t}_1$, $\tilde{t}_2$ and $\tilde{t}_3$, and are therefore considered as consistently capturing the correct prior. We shall treat the composite tests as a baseline to study the effect of prior misspecification, by comparing them to tests using a single $\tilde{t}$.  }

For each simulation configuration considered in this study, the proposed procedures are as powerful as the procedure with the best performance, except for the cases when $B$ is sparse, $p$ is small, and priors are severely misspecified in the proposed tests; {see Figure \ref{fig:SA_Sigma_toep_k3_t001_LR} in the Supplementary Material}. We highlight the following observations based on the simulation results. 

%\vspace{-.1cm}
\begin{itemize}\itemsep-.2ex
\item[(1)] {The composite tests are slightly less efficient than BNP$_{\rm ridge}$ and BNP$_{\rm high}$ when the correct prior $\tilde{t}_1$ is used, as in Figure \ref{fig:SA_Sigma_dense_k5_t100_BNP}. However, as in Figure \ref{fig:SA_Sigma_dense_k5_t001_LH}, when the prior is severely misspecified, the composite test is significantly more powerful. It suggests that the composite tests are robust against prior misspecification, although losing some efficiency against tests with correctly specified priors.}
% Figure 3
\begin{figure}[htbp]
\setlength{\belowcaptionskip}{-5pt}
\includegraphics[width = \linewidth,height = 0.4\linewidth]{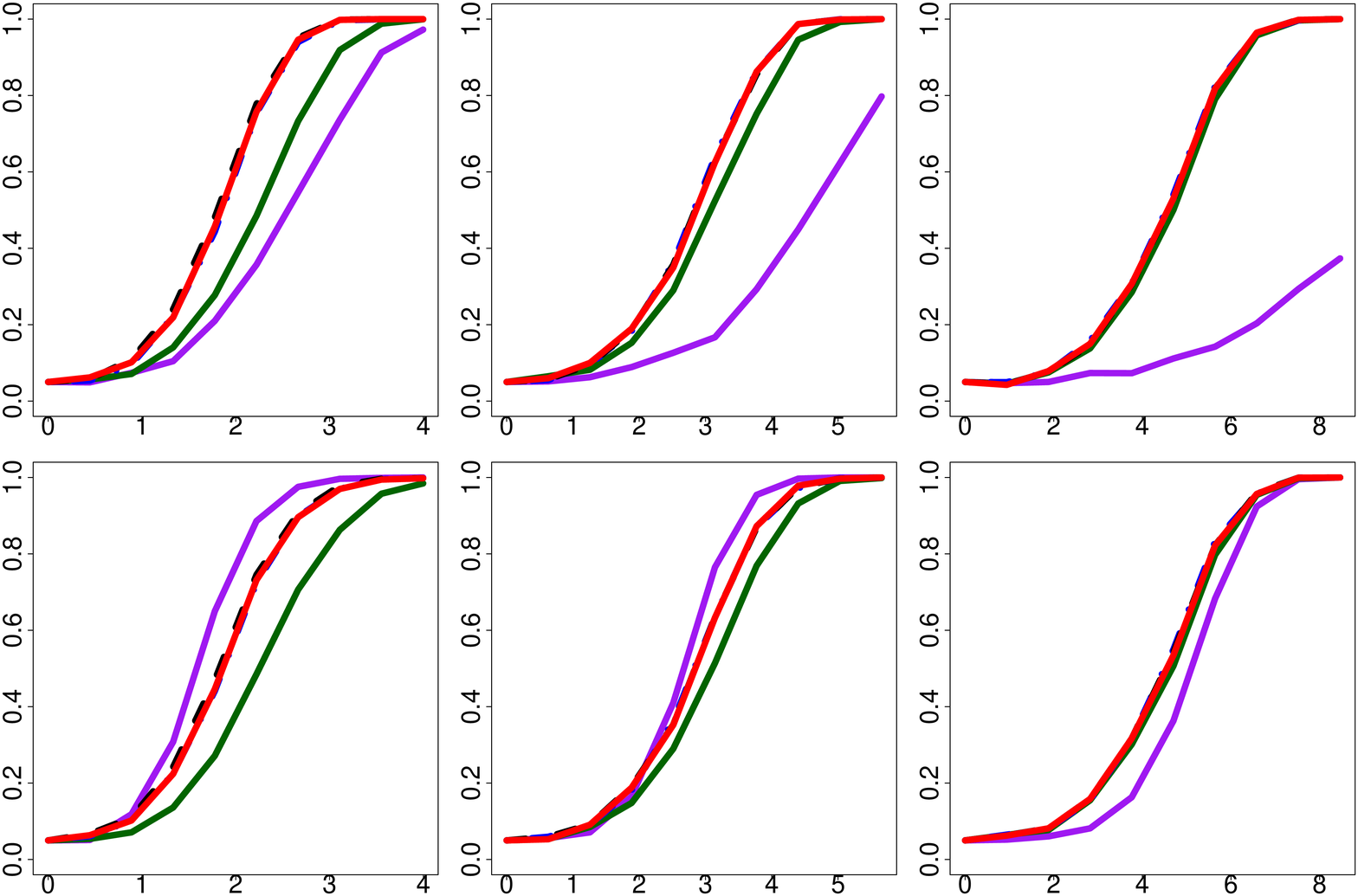}
\caption{\footnotesize Size-adjusted power with $\Sigma=\Sigma_{toep}$,~$k=3$. Rows (top to bottom): $B =\mbox{Dense and Sparse} $; Columns (left to right): $p = 150, 600, 3000$. LR$_{\rm comp}$ (red, solid);  ZGZ (green, solid); oracle CX (purple, solid); LR$_{\rm ridge}$ (black, dashed) and LR$_{\rm high}$ (blue, dotted-dashed) with $\tilde{t} = (0,1,0)$.}
\label{fig:SA_Sigma_toep_k3_t010_LR}
\end{figure}
\item[(2)] {Although ridge-shrinkage and higher-order shrinkage behave similarly under the correct prior, the latter outperforms the former when the prior is misspecified; see Figure \ref{fig:SA_Sigma_dense_k5_t001_LH}. This provides evidence for the robustness of high-order shrinkage against unfavorable ridge shrinkage parameter selection. } 

%They all outperform ZGZ, especially for $p =150$, $600$ and if eigenvalues of $\Sigma_p$ are disperse; see Figure \ref{fig:SA_Sigma_toep_k3_t001_LR}. The composite tests are slightly less efficient.
\item[(3)] ZGZ is a special case of the proposed test family with $f(x)=1$ for all $x$, which amounts to replacing $\hSigma$ with $I_p$. When $\Sigma_p=I_p$, ZGZ appears to be the reasonable option at least intuitively. Note, both $\frakF_{\rm ridge}$ and $\frakF_{{\rm high}}$ contain functions close to $f(x)=1$. Figures for $\Sigma_p = I_p$ displayed in Section \ref{sec:additional_simulation} of the Supplementary Material show that the proposed tests perform as well as ZGZ in that case.  It may be viewed as evidence of the effectiveness of the data-driven shrinkage selection strategy detailed in Section \ref{sec:locally_most_powerful_test}.

\item[(4)] {Comparing to ZGZ, when the eigenvalues of $\Sigma_p$ are disperse, the proposed tests are significantly more powerful when $p=150$ and $600$, but behave similarly as ZGZ when $p=3000$. On the other hand, as in Figure \ref{fig:SA_Sigma_dense_k5_t001_LH}, the ridge-regularized test with a severely misspecified prior $\tilde{t}_3$, is close to ZGZ. }

%\item[(3)] When $\tilde{t} = (0,0,1)$, ridge-regularized tests are very close to ZGZ. However, Figure \ref{fig:SA_Sigma_dense_k5_t001_LH} shows the advantage of higher-order regularized tests over ZGZ and ridge regularization. This provides evidence for the robustness of high-order shrinkage against unfavorable ridge shrinkage parameter selection.

\item[(5)] CX is a test specifically designed for sparse alternatives. The procedure shows its advantage in favorable settings, especially when $p=150$. Simulation results suggest that the proposed tests are still comparable to CX even under sparse $BC$ and $\Sigma_p^{-1}$, as long as the prior in use is not severely misspecified. When $p$ is large, the proposed tests are significantly better when $\Sigma_p = I_p$. Evidence may be found in Figures \ref{fig:SA_Sigma_ID_k3_t100_LH},  \ref{fig:SA_Sigma_ID_k5_t010_BNP} and \ref{fig:SA_Sigma_ID_k3_t001_LR} of the Supplementary Material. 
\end{itemize}

\bibliographystyle{apalike}
\bibliography{GLHT}

%%%%%%%%%%%
%% Appendix
%%%%%%%%%%%

\setcounter{section}{0}
\renewcommand{\thesection}{A.\arabic{section}}
\setcounter{equation}{0}
\renewcommand{\theequation}{A.\arabic{equation}}
\setcounter{subsection}{0}
\renewcommand{\thesubsection}{A.\arabic{subsection}}
\setcounter{table}{0}
\renewcommand{\thetable}{A.\arabic{table}}
\setcounter{figure}{0}
\renewcommand{\thefigure}{A.\arabic{figure}}

\setcounter{theorem}{0}{}
\renewcommand\thetheorem{A.\arabic{theorem}}
\setcounter{lemma}{0}
\renewcommand\thelemma{A.\arabic{lemma}}
\setcounter{proposition}{0}
\renewcommand\theproposition{A.\arabic{proposition}}

\section*{Appendix: Proof of Theorem \ref{thm:asmp_null}}
%\label{subsec:proof_Theorem_asmp_null}
This appendix contains a proof outline of Theorem \ref{thm:asmp_null}. Additional proofs of supporting lemmas and other theorems can be found in the Supplementary Material. %The arguments used here build on those in \citet{bai2010spectral} and \citet{pan2011central} but contain considerable innovations. First, the key assumption \textbf{C4} on the spectrum of $\Sigma_p$ is different from its counterpart in \citet{bai2010spectral}. Therefore, additional results regarding the Stieltjes transforms $m_{n,p}(\comp)$ and $m(\comp)$ are needed here. We also need the uniform convergence of derivatives of $m_{n,p}(\comp)$ on a contour, which was not part of \citet{bai2010spectral}. These results are detailed in Section \ref{sec:m_z_on_C} of the Supplementary Material. Second, \citet{pan2011central} only considered the case $\Sigma_p=I_p$ because of the invariance of Hotelling's $T^2$ statistic with respect to $\Sigma_p$. Important arguments in their paper, for example their Lemma 6, no longer hold under general covariance structures. Additionally, the calculation of the asymptotic variance and covariance of the quadratic form $n^{-1}Q_n^T \bY^T (\hSigma-\comp I_p)^{-1}\bY Q_n$ is significantly more involved than for Hotelling's $T^2$ statistic. Third, our proof includes an important transformation of the quadratic form to deal with the complex correlation structure between $\bY Q_n$ and $\hSigma$. We believe the trick can be used to generalize existing work in the literature, for example \citet{bai2017limiting}.

%\subsection{Proof of Theorem \ref{thm:asmp_null}}

%In the proof, we frequently refer to \cite{PanZhou2011}. 
Recall that $Q_n= X^T (XX^T)^{-1} C[C^T (XX^T)^{-1}C]^{-1/2}.$ 
%\begin{align*}
%Q_n    &= X^T (XX^T)^{-1} C[C^T (XX^T)^{-1}C]^{-1/2},\\
%\bM(f)   &= \frac{1}{n} Q_n^T \bY^T f(\hSigma)\bY Q_n,\\
%\hSigma&= \frac{1}{n} \bY(I- X^T(XX^T)^{-1}X) \bY^T. 
%\end{align*}
Introduce the product $Q_n = U_nV_n$ with
\begin{align}
U_n &= X^T (XX^T)^{-1/2}~,\label{eq:U_n} \\
V_n &= (XX^T)^{-1/2} C [C^T(XX^T)^{-1}C]^{-1/2}~.\label{eq:V_n}
\end{align}
This decomposition will aid the analysis of the correlation between $\bY Q_n$ and $\hSigma$.

From now on, use  $\Sigma_{p}^{T/2}$ to denote $(\Sigma_p^{1/2})^T$. %and we shall use $\bz_j$ to denote the $j$th column of $\bZ$. 
 Under the null hypothesis, the following representations hold: 
\begin{align*}
\bM(f)    &= \frac{1}{n} V_n^T U_n^T \bZ^T \Sigma_p^{T/2}f(\hSigma) \Sigma_p^{1/2}\bZ U_n V_n,\\ 
\hSigma &= \frac{1}{n} \Sigma_p^{1/2}\bZ (I-U_n U_n^T) \bZ^T \Sigma_p^{T/2}. 
\end{align*}
%where we invent the notation $\Sigma_{p}^{T/2}$ to mean $(\Sigma_p^{1/2})^T$. From now on, we use $\bz_j$ denotes the $j$th column of $\bZ$. 
%
%To avoid potential ambiguity, from now on, we set the space of random elements to be the class of all continuous functions $h(\comp) = [h(\comp)]_{i,j=1}^k: \comp\in\calC \to \mathbb{C}^{k^2}$ equipped with the metric: 
%\[\rho(h(\comp), g(\comp)) = \sup\limits_{\comp\in\calC} \sum\limits_{i,j=1}^k \Big|[h(\comp)]_{i,j}-[g(\comp)]_{i,j}\Big|,\]
%with the convention that a lower dimensional random element is also considered belonging to the space with 0's filled in on the deficient coordinates. 
%
 %given the metric 
%\[\mbox{Distance}(h(\comp), g(\comp)) = \sup\limits_{\comp\in\calC} \|h(\comp)-g(\comp)\|_1,\] 
%for any two points $h(\comp)$ and $g(\comp)$ within the space. 
%
Observe that the joint asymptotic normality of entries in $\sqrt{n}\bM(f)$ is equivalent to the asymptotic normality of
\[n^{-1/2} \alpha^T V_n^T U_n^T \bZ^T \Sigma_p^{T/2}f(\hSigma)\Sigma_p^{1/2} \bZ U_n V_n \eta\]
for arbitrary (but fixed) vectors $\alpha$ and $\eta\in\mathbb{R}^q$ . 

Recall that $\calX = [0,~\limsup_{p}\lambda_{\max}(\Sigma_p)(1+\sqrt{\gamma})^2]$. Let $\calC$ be any contour enclosing $\calX$ such that $f(\cdot)$ is analytic on its interior. With slight modifications, all arguments in the following hold for arbitrary such $\calC$. For convenience, select $\calC$ as rectangle with vertices $\underline{u}\pm iv_0$ and $\overline{u}\pm iv_0$, such that $v_0>0;~ \overline{u} > \limsup\lambda_{\max}(\Sigma_p)(1+\sqrt{\gamma})^2;~ \underline{u}<0.$ 
Such a rectangle must exist. 

By \emph{Cauchy's integral formula}, if $\lambda_{\max}(\hSigma)<\overline{u}$,
\begin{equation}
\label{eq:tansf_cauchy}
\begin{split}
n^{-1/2}&\alpha^T V_n^T U_n^T \bZ^T\Sigma_p^{T/2} f(\hSigma)\Sigma_p^{1/2} \bZ U_n V_n \eta \\
&=\frac{-1}{2\pi i} \oint_\calC f(\comp) n^{-1/2} \alpha^T V_n^T U_n^T \bZ^T \Sigma_p^{T/2}(\hSigma-\comp I)^{-1}\Sigma_p^{1/2} \bZ U_n V_n \eta d\comp. 
\end{split}
\end{equation}
If $\lambda_{\max}(\hSigma)\geq \overline{u}$, the above equality may not hold. However, if we can show that
$\mP(\lambda_{\max}(\hSigma)\geq\overline{u})$ converges to 0, we can still acquire the weak limit of the left-hand side by deriving the weak limit of the right-hand side.
\citet[Theorem 3.1]{yin1988limit} implies that 
\begin{equation}
\label{eq:largest_eigen_hSigma_not_exceed_u}
\mP(\lambda_{\max}(\hSigma)\geq\overline{u}) \to 0.
\end{equation} 
Hence, it suffices to show the asymptotic normality of the process % $\xi_{n}(\comp,\alpha,\eta)$. 
\[
\xi_{n}(\comp,\alpha,\eta) = n^{-1/2}\alpha^T V_n^T U_n^T \bZ^T \Sigma_p^{T/2} (\hSigma-\comp I)^{-1}\Sigma_p^{1/2} \bZ U_n V_n \eta, \qquad \comp \in\calC.
\]
Clearly, $\xi(\comp,\alpha,\eta)$ is continuous with respect to $\comp$. All asymptotic results are derived in the space of continuous functions on $\calC$ with uniform topology. %In the following, study the $k\times k$ matrix process $U_n^T \bZ^T \Sigma_p^{T/2} (\hSigma-\comp I)^{-1}\Sigma_p^{1/2} \bZ U_n$, that is a component of $\xi(\comp,\alpha,\eta)$.  Note that a $k\times k$ complex-valued matrix can be viewed as a $2k^2$-variate real-valued random element. To unify dimensionality and avoid potential ambiguity, all objects are treated as random elements in the metric space $C(\calC,\mR^{2k^2})$ in the following. For dealing with matrices of dimension $q\times q$ (with $q < k$), note that a $q\times q$ matrix can be viewed as the $q$-th order leading submatrix of a $k\times k$ matrix whose other entries equal 0. Similarly, univariate functions may be viewed as $1\times 1$ submatrices. Therefore, 
Results in Chapter 2 of \citet{billingsley1968convergence} apply with Euclidean distance replaced by Frobenius norm of a matrix, that is $\|A\|_F = (\sum_{i=1}^m \sum_{j=1}^r |a_{ij}|^2)^{1/2}$, where $A= [a_{ij}]_{ij}$.

We may proceed to prove the asymptotic normality of $\xi_n(\comp,\alpha,\eta)$ on $\comp\in\calC$ directly. However, several technical challenges need to be addressed. First, in view of the spectral norm of $(\hSigma-\comp I)^{-1}$ being unbounded when $\comp$ is close to the real axis and extreme eigenvalues of $\hSigma$ exceed $\limsup\lambda_{\max}(\Sigma_p)(1+\sqrt{\gamma})^2$, the tightness of the process $\xi_{n}(\comp,\alpha,\eta)$ is unclear. Secondly, $\hSigma$ is not a summation of independent terms, but contains $\bZ U_n U_n^T \bZ^T$, a component containing cross product terms between pairs of columns of $\bZ$. These terms entangle the analysis of the correlation between $\hSigma$ and each single column of $\bZ$. %make the correlation between $\hSigma$ and each column of $\bZ$ complicated. 
For these technical reasons, we avoid directly working on $\xi_{n}(\comp,\alpha,\eta)$ under \textbf{C1} on $\comp\in\calC$, but start with $n^{-1/2}U_n^T \bZ^T \Sigma_p^{T/2}(\tSigma-\comp I)^{-1}\Sigma^{1/2}_p \bZ U_n$, a component of $\xi_{n}(\comp,\alpha,\eta)$ with $\hSigma$ replaced by an uncentered counterpart
\begin{equation}
\tSigma = \frac{1}{n} \Sigma_p^{1/2}\bZ\bZ^T\Sigma_p^{T/2}.
\label{eq:hSigma_tSigma_idenity}
\end{equation}
The relationship between $\tSigma$ and $\hSigma$ is given by
\begin{equation}\label{eq:tSigma_hSigma_identity}
\hSigma = \tSigma -\frac{1}{n} \Sigma_p^{1/2} \bZ U_n U_n^T \bZ^T\Sigma_p^{T/2}.
\end{equation}
Next, we modify the process and the distribution of $\bZ$ as follows.

\emph{Process smoothing.}\hspace{0.1 cm} Select a sequence of positive numbers $\rho_n$ decaying to 0 with a rate such that for some $\omega\in(1,2)$,
\[ n \rho_n \downarrow 0, \quad \rho_n \geq n^{-\omega}. \]  
Let $\calC^+ = \calC \cap \{ u + iv\colon |v|\geq \rho_n  \}$.  Define
\begin{align*}
&\tcalQ_n(\comp) = n^{-1}  U_n^T \bZ^T \Sigma_p^{T/2}(\tSigma-\comp I)^{-1}\Sigma_p^{1/2} \bZ U_n, \quad \mbox{ if }\comp\in\calC^+ ,\\
&\tcalQ_n(\comp) = \frac{\rho_n-v}{2\rho_n}\tcalQ_n(u + i \rho_n) + \frac{v+\rho_n}{2\rho_n} \tcalQ_n(u - i \rho_n), \quad \mbox{ if } \comp\in\calC\setminus\calC^+.
\end{align*}
To understand this definition better, note that if $\comp$ is too close to the real axis,  $\tcalQ_n(\comp)$ is modified to be the linear interpolation of its values at $u + i \rho_n$ and $u - i \rho_n$. Observe that $V_n$ appearing in $\xi_n(\comp,\alpha,\eta)$ was left out when defining $\tcalQ_n(\comp)$. This trick that helps transforming back to $\hSigma$ from $\tSigma$; see \eqref{eq:tQ_hQ}. Note also that $V_n$ is a sequence of deterministic matrices of fixed dimensions, having a limit under \textbf{C5} and \textbf{C6}.  %The reason to start with $\tSigma$ is that it can be written as a sum of independent terms, which are the outer product of each column of $\bZ$ with itself. 
The reason to smooth the process is to guarantee a bound of order $O(\rho_n^{-1})$ on the spectral norm of $(\tSigma-\comp I_p)^{-1}$. It is crucial in the proof of tightness. %We will first show asymptotics of $\tcalQ(\comp)$ and later transform back to the integral of $\xi(\comp,\alpha,\eta)$ on $\calC$. 

\emph{Variable truncation.}\hspace{0.1cm} \textbf{C1} will be temporarily replaced by the following truncated variable condition. Select a positive sequence $\varepsilon_n$ such that 
\begin{equation*}
\varepsilon_n \to 0 \qquad \mbox{and} \qquad \varepsilon_n^{-4} \mE[ z_{11}^4\mathbbm{1}(|z_{11}|\geq \varepsilon_nn^{1/2})]\to0.
\end{equation*}
The existence of such a sequence is shown in \citet{yin1988limit}. We then truncate $z_{ij}$ to be $z_{ij}\mathbbm{1}(|z_{ij}|\leq\varepsilon_nn^{1/2})$. After that, we re-standardize the truncated variable to maintain zero mean and unit variance. Since we will mostly work on the truncated variables in the following sections, for notational simplicity, we shall use $z_{ij}$ to denote the truncated random variables and $\breve{z}_{ij}$ to denote the original random variable satisfying \textbf{C1}. That is,
\[
z_{ij} = \frac{ \breve{z}_{ij}\mathbbm{1}(|\breve{z}_{ij}| \leq \varepsilon_n n^{1/2}) - \mE \breve{z}_{ij}\mathbbm{1}(|\breve{z}_{ij}| \leq \varepsilon_n n^{1/2})}{ \{\mE [\breve{z}_{ij}\mathbbm{1}(|\breve{z}_{ij}| \leq \varepsilon_n n^{1/2}) - \mE \breve{z}_{ij}\mathbbm{1}(|\breve{z}_{ij}| \leq \varepsilon_n n^{1/2})]^2\}^{1/2}}.
\]
For some constant $\calK$, when $n$ is sufficiently large, 
\begin{equation} 
|z_{ij}| \leq \calK \varepsilon_n n^{1/2}, \quad \mE [z_{ij}]=0, \quad \mE [z_{ij}^2] = 1, \quad \mE[ z_{ij}^4] <\infty.
\label{eq:truncated_variable_condition}
\end{equation}

The reason to truncate $\breve{z}_{ij}$ is to obtain a bound on the probability of extreme eigenvalues of $\hSigma$ exceeding $\limsup_p \lambda_{\max}(\Sigma_p)(1+\sqrt{\gamma})^2$. A tail bound decaying fast enough is critical when proving tightness of the smoothed random processes on $\calC$. Under the original condition \textbf{C1}, although \eqref{eq:largest_eigen_hSigma_not_exceed_u} holds, such a tail bound is not available. After the truncation, the following lemma shown in \citet{yin1988limit,bai2004clt} holds. 
\begin{lemma}
\label{lemma:poly_bound_tSigma}Suppose the entries of $\bZ$ satisfy \eqref{eq:truncated_variable_condition}. For any positive $\ell$ and any $\mathfrak{D}$ such that $\mathfrak{D} \in (\limsup_p\lambda_{\max}(\Sigma_p)(1+\sqrt{\gamma})^2, ~\overline{u})$,
\[
\mP ( \lambda_{\max}(\tSigma) \geq \mathfrak{D} ) = o(n^{-\ell}).
\]
%If $\gamma < 1$ and $\liminf_n \lambda_{\min}(\Sigma_p) > 0$, for any $\eta_2 \in (\underline{u}, ~\liminf_n \lambda_{\min}(\Sigma_p) (1-\sqrt\gamma)^2)$, 
%\[\mP ( \lambda_{\min}(\tSigma) \leq \eta_2 ) = o(n^{-\ell}).\]
\end{lemma}
%It's worth mentioning that even without the variable truncation, extreme eigenvalues of $\tSigma$, also $\hSigma$, are asymptotically smaller than $\mathfrak{D}$ almost surely, as $n,p\to\infty$ \citep{yin1988limit}. However, we won't achieve such a tail bound decaying faster than any polynomial of $n$. %If further $\bZ$ is sub-gaussian, an exponential bound of the above probability is known in literature \citep{vershynin2010introduction}. 
%
%From now on, all objects introduced before are re-defined on $z_{ij}$ instead of $\breve{z}_{ij}$. All results in the rest of the section are constructed with respect to $z_{ij}$. 
%
It is argued later that the process smoothing and variable truncation steps do not change the weak limit of objects under consideration. 

\begin{theorem}
\label{thm:convergence_Gabz}
For arbitrary vectors $a$ and $b\in\mathbb{R}^k$, define $G_n(\comp, a, b) = a^T \tcalQ_n(\comp)b$. Suppose $\bZ$ satisfies  (\ref{eq:truncated_variable_condition}) and suppose \textbf{C2}--\textbf{C6} in Section \ref{sec:methodology} hold. Then,
\[n^{1/2} \Big\{ G_n(\comp, a,b)  -  a^Tb~ \frac{\Theta(\comp,\gamma)-1}{\Theta(\comp,\gamma)} \Big\}  \stackrel{D}{\longrightarrow} {\Psi}^{(1)}(\comp), \qquad \comp\in \calC,\]
where $\stackrel{D}{\longrightarrow}$ denotes weak convergence in $C(\calC, \mathbb{R}^{2})$, and ${\Psi}^{(1)}(\comp)$ is a Gaussian process with zero mean and covariance function %$\bGamma^{(1)}(\comp_1, \comp_2)$ defined as 
\begin{align*}
%&\bGamma(\comp_1,\comp_1) = \frac{\gamma\Theta_2(\comp,\gamma)}{(1+\gamma\Theta_1(\comp,\gamma))^4}[\|\alpha\|^2\|\beta\|^2 + (\alpha^T\beta)^2].
&\Gamma^{(1)}(\comp_1,\comp_2) = {\delta(\comp_1,\comp_2,\gamma)}\Theta^{-2}(\comp_1,\gamma)\Theta^{-2}(\comp_2,\gamma)[\|a\|^2\|b\|^2 + (a^Tb)^2].
\end{align*}
\end{theorem}
See Section \ref{sec:proof_Theorem_Gabz} of the Supplementary Material for proof of the theorem.

Next, transforming back to $\hSigma$, define 
\begin{align*}
\hcalQ_n(\comp) &= n^{-1} U_n^T \bZ^T \Sigma_p^{T/2}(\hSigma-\comp I)^{-1}\Sigma_p^{1/2} \bZ U_n, \qquad\comp\in\calC^+, \\
\hcalQ_n(\comp) &= \frac{\rho_n-v}{2\rho_n}\hcalQ_n(u + i \rho_n) + \frac{v+\rho_n}{2\rho_n} \hcalQ_n(u - i \rho_n), \qquad \comp\in\calC\setminus\calC^+.
\end{align*}
Using the identity (\ref{eq:hSigma_tSigma_idenity}),
and Lemma \ref{lemma:Woodbury_formula} (\emph{Woodbury matrix identity}) in the Supplementary Material, we get
\begin{equation}
\label{eq:tQ_hQ}
\hcalQ_n(\comp) = \tcalQ_n(\comp) [I_k- \tcalQ_n(\comp)]^{-1}.
\end{equation}
Notably, 
$(\Theta(\comp,\gamma)-1)/\Theta(\comp,\gamma)$ is bounded away from 1 on $\calC$. Since $\hcalQ_n(\comp)$ is a smooth function of $\tcalQ_n(\comp)$, applying the delta-method, the following result is an immediate consequence of Theorem \ref{thm:convergence_Gabz}.
\begin{lemma}
\label{lemma:hat_Q_n_converge}
Suppose $\bZ$ satisfies  (\ref{eq:truncated_variable_condition}) and suppose \textbf{C2}--\textbf{C6} in Section \ref{sec:methodology} hold. Then, 
\[n^{1/2} \{ \hcalQ_n(\comp) - \{\Theta(\comp,\gamma)-1\}I_k  \} \stackrel{D}{\longrightarrow}  \Psi^{(2)}(\comp),\qquad \comp\in\calC, \]
where $\stackrel{D}{\longrightarrow}$ denotes weak convergence in $C(\calC, \mathbb{R}^{2k^2})$, and $\Psi^{(2)}(\comp)=[\Psi^{(2)}(\comp)]_{ij}$ is a $k\times k $ symmetric Gaussian matrix process with zero mean and covariance, such that for $i\leq j$, $i'\leq j'$,
\begin{align*}
&\mE[\Psi^{(2)}(\comp_1)]_{ii}[\Psi^{(2)}(\comp_2)]_{ii} =  {2\delta(\comp_1,\comp_2,\gamma)},\\
&\mE[\Psi^{(2)}(\comp_1)]_{ij}[\Psi^{(2)}(\comp_2)]_{ij} = {\delta(\comp_1,\comp_2,\gamma)} , \quad \mbox{ if } i\neq j, \\
&\mE[\Psi^{(2)}(\comp_1)]_{ij}[\Psi^{(2)}(\comp_2)]_{i'j'} =0, \quad \mbox{ if } i\neq i' \mbox{ or }j\neq j'. 
\end{align*}
\end{lemma}

 Define a smoothed version of $\xi_n(\comp,\alpha,\eta)$ as %The calculation of the covariance kernel of $\Psi^{(3)}(\comp)$ is basic, though tedious, and hence details are omitted.
%To smooth $\xi_{n}(\comp,\alpha,\eta)$ in the same way as $\hcalQ_n(\comp)$, define
\begin{align*}
\widehat{\xi}_n(\comp,\alpha,\eta) &= 
  \xi_n(\comp,\alpha,\eta), &\comp\in\calC^+ ,& \\
\widehat{\xi}_n(\comp,\alpha,\eta)&=  \frac{\rho_n-v}{2\rho_n} {\xi_n}( u + i \rho_n,\alpha,\eta) +  \frac{v+\rho_n}{2\rho_n}\xi_n( u - i \rho_n,\alpha,\eta), & \comp\in\calC\setminus\calC^+.
  %\label{eq:trunct_tM}
\end{align*}
%Note that $\widehat{\xi}_n(\comp,\alpha,\eta) = n^{1/2} \alpha^T V_n^T \hcalQ_n(\comp) V_n\eta$ is a smooth function of $\tcalQ_n(\comp)$. 
We immediately have the following lemma. 
\begin{lemma}
\label{lemma:hat_xi_converge}
Suppose that $\bZ$ satisfies  (\ref{eq:truncated_variable_condition}) and \textbf{C2}--\textbf{C6} hold. Then,
%in Section \ref{sec:methodology} holds, 
\begin{equation*}
\widehat{\xi}_n(\comp,\alpha,\eta) - n^{1/2} (\Theta(\comp,\gamma)-1)\alpha^T\eta \stackrel{D}{\longrightarrow }\Psi^{(3)}(\comp),
\end{equation*}
where $\stackrel{D}{\longrightarrow}$ denotes weak convergence in $C(\calC, \mathbb{R}^{2})$, and ${\Psi}^{(3)}(\comp)$ is a Gaussian process with zero mean and covariance function
\begin{equation*}
\Gamma^{(2)}(\comp_1,\comp_2) = {\delta(\comp_1,\comp_2,\gamma)}[\|\alpha\|^2\|\eta\|^2 + (\alpha^T\eta)^2].
\end{equation*}
\end{lemma}
%
%It is known in literature that $\Theta(\comp,\gamma)$ is bounded on $\calC$. The process smoothing  Hence, 
%\begin{align*}
%&n^{1/2}\oint_{\calC\setminus\calC^+} f(\comp) [\Theta(\comp,\gamma)-1] d\comp \to0,\\
%&\oint_{\calC^2\setminus(\calC^+)^2} f(\comp_1)f(\comp_2) \delta(\comp_1,\comp_2,\gamma) d\comp_1d\comp_2 \to0,
%\end{align*}
%
The following result is an immediate consequence of the foregoing:
%recalling notation in Theorem \ref{thm:asmp_null},
\begin{align*}
\oint_\calC \frac{f(\comp)\widehat{\xi}_n(\comp,\alpha,\eta)}{-2\pi i}d\comp -n^{1/2}\Omega(f,\gamma) \alpha^T\eta  {\Longrightarrow } \mathcal{N}(0,{[\|\alpha\|^2\|\eta\|^2 + (\alpha^T\eta)^2]\Delta(f,\gamma)}). 
\end{align*}
In Section \ref{sec:truncation} of the Supplementary Material (see Lemma \ref{lemma:truncation_not_change_local_power_Mf} and (\ref{eq:integral_negligible_near_x_axis}) for details), we  verify that, if we replace $\widehat{\xi}_n(\comp,\alpha,\eta)$ with $\xi_n(\comp,\alpha,\eta)$, and (\ref{eq:truncated_variable_condition}) with \textbf{C1}, the above result continues to hold.

%Since $\sqrt{n}\alpha^T \bM(f)\eta=(-2\pi i)^{-1} \oint_\calC {f(\comp){\xi}_n(\comp,\alpha,\eta)}d\comp$ when $\lambda_{\max}(\hSigma)<\overline{u}$ and (\ref{eq:largest_eigen_hSigma_not_exceed_u}) holds, the proof of Theorem \ref{thm:asmp_null} is complete.

\section*{Supplementary Material}
\label{SM}
Supplementary Material includes additional simulation results and detailed proofs of the main theoretical results presented in this
paper.

\end{document}